\documentclass[twocolumn,aps,prc,amsmath,amssymb,superscriptaddress]{revtex4-2}
\usepackage[dvips]{graphicx}
\usepackage{epsfig}
\usepackage{amsmath}
\usepackage[dvips]{color}
\newcommand{\vis}[1]{\mbox{\boldmath $#1$}}
\usepackage{bm}
\usepackage{xcolor}
\begin{document}

\title{Three-body model of $^{6}$He with non-local halo effective field theory potentials}
\author{E. C. Pinilla}
\affiliation{Universidad Nacional de Colombia, Sede Bogot\'a, Facultad de Ciencias, Departamento de F\'\i{}sica, Grupo de F\'\i{}sica Nuclear, Carrera 45 N\textsuperscript{o} 26-85, Edificio Uriel Guti\'errez, Bogot\'a D.C. C.P. 1101, Colombia.}
\affiliation{Department of Physics, University of Trento, V. Sommarive 14, I-38123 Trento, Italy}
\affiliation{INFN-TIFPA, V. Sommarive 14, I-38123 Trento, Italy}
\author{W. Leidemann}
\author{G.  Orlandini}
\affiliation{Department of Physics, University of Trento, V. Sommarive 14, I-38123 Trento, Italy}
\affiliation{INFN-TIFPA, V. Sommarive 14, I-38123 Trento, Italy}
\author{P. Descouvemont}
\affiliation{Département de Physique, C.P. 229,\\
Universit\'e libre de Bruxelles (ULB), B 1050 Brussels, Belgium}
\date{\today}

\begin{abstract}
We study the $^6$He Borromean nucleus in coordinate representation within a three-body model with two-body potentials derived from cluster effective field theory (EFT). These potentials are originally developed in momentum space and Fourier transformed to provide non-local potentials in configuration space.  We use hyperspherical coordinates in combination with the Lagrange-mesh technique to compute the ground state energy, root mean square radius and the E1 strength distribution of $^6$He.  We also introduce a three-body interaction to eliminate dependencies on the cutoff parameter of the two-body potentials on the ground state energy.  The E1 strength distribution exhibits a low lying resonance. However it is strongly influenced by the choice of the three-body EFT interaction. 
\end{abstract}

\maketitle

\section{Introduction}
Modern nuclear theory is paying a lot of attention to develop state of the art models as precise as possible to study nuclei. In particular, the derivation of nucleus-nucleus interactions from ab initio theories that allow the prediction rather than the phenomenological explanation of nuclear processes is of current interest.

Effective field theory (EFT) is a powerful framework that takes advantage of the separation of scales in a physical system by integrating out its irrelevant degrees of freedom \cite{We79,We80}.  In nuclear physics, EFTs have become very popular to study few nucleon configurations (see for instance Refs.  \cite{BK02,HJP17,HKK20,MEF21}).

An interesting application of the EFT framework  is the study of halo nuclei. Those nuclei can be seen as made up of a core plus one or more nucleons that are weakly bound to the core \cite{Al04}. Thus, halo nuclei possess a separation of scales: a low momentum scale associated with the weakly bound character of the halo, and a high-momentum scale that corresponds to the binding energy of the core. 

A typical example of halo nuclei is $^6$He, which can be described as an alpha core plus two weakly bound neutrons forming the halo.  This nucleus is called Borromean,  since none of the two-body pairs $\alpha-n$ or $n-n$ form a bound state. Three-body models with phenomenological local potentials have been widely applied to study the structure and dynamics of this nucleus \cite{ZDF93,DTV98,DDB03,BCD09,RAG08}.

In Ref.\ \cite{JEP14}, a cluster EFT was used to find the ground state energy of the nucleus $^6$He. This is a leading order (LO) three-body halo EFT and assumes that the two-body interactions are dominated by the $S_{0}$ $n-n$ and $P_{3/2}$ $\alpha-n$ partial waves. Other partial waves are regarded as higher-order corrections in their EFT.  A three-body EFT interaction was derived at LO to accurately reproduce the $^6$He ground state energy and to eliminate cutoff dependencies. Once the two-body potentials are fixed, the three-body problem is solved using momentum-space Faddeev equations \cite{Fa61}. 

Cluster EFT potentials are derived in a natural way in momentum space. However,working in this space leads to convergence problems when the Coulomb interaction is present.Thus, a treatment in configuration representation is more convenient, since the Coulomb interaction can be managed in a simpler way.   In coordinate space,  those potentials become non-local and can be obtained by performing a double Fourier transform.  

In principle, the cluster EFT potentials \cite{FA20,F22,CFJ24} are more fundamental than the phenomenological ones, since they are not fitted to reproduce experimental data. Instead, they are derived from the matching to a low-energy theory of the expansion of an observable in a series, up to a chosen order, of the ratio of a typical low-momentum scale, over a high momentum scale.  In our case, the low-energy theory is the effective range theory.  The low energy constants (LECs) are related to terms of the expansion, which in turn, are related to the effective range parameters.  Once, the experimental values of these parameters are inserted, we can get the cluster EFT potentials at some order. 

In this work, we employ two-body cluster EFT non-local potentials to study $^6$He.  We use a three-body model in configuration space with hyperspherical coordinates and the Lagrange-mesh technique \cite{TBD05}. In particular, we compute structure properties such as the ground state energy,  the root mean square radius (rms), as well as the electric dipole strength distribution.  We Fourier transform the cluster EFT $S$-wave $n-n$ potential, and the $s_{1/2}$ and $p_{3/2}$ $\alpha-n$ potentials derived in Refs.  \cite{FA20,F22,CFJ24} at LO.  An $S$-wave EFT three-body potential in representation coordinates at LO is introduced aiming at getting the correct binding energy of $^6$He and to eliminate cutoff dependencies on the two-body potentials. 

The constructed EFT non-local $\alpha-n$ potential binds a bound state in the $S$-wave that simulates a forbidden state.  As this state gives spurious eigenvalues in the three-body Hamiltonian, we must remove it. In this paper, we address the elimination of forbidden states in cluster EFT potentials by extending the projection technique given in Ref.\ \cite{KP78} to non-local potentials.

The paper is organized as follows.  In sec. \ref{sec_3B}, we describe the three-body model with non-local potentials in hyperspherical coordinates. Section \ref{sec_2BEFT} summarizes the derivation of the two-body potentials and we describe how to remove the forbidden states. In Sec. \ref{sec_E1}, we describe the calculation of the dipole strength distribution. Section \ref{sec_6he} shows the application to $^6$He.  Concluding remarks are given in Sec.  \ref{sec_conclu}.
\section{Three-body model with non-local potentials}
\label{sec_3B}
\subsection{Coupled differential equations in hyperspherical coordinates}
Let us consider a three-body nucleus, made of three clusters, each with nucleon numbers $A_i$ and position coordinates $\vis{r}_i$. The Hamiltonian of the system, for two-cluster interactions only, is given by
\begin{equation}
H=\sum\limits_{i=1}^{3}\frac{\vis{p^2_i}}{2m_NA_i}+\sum\limits_{i<j=1}^{3}V_{ij},
\label{Ham3b}
\end{equation}
with $m_N$ the nucleon mass and $\vis{p}_i$ the momentum of cluster $i$.  

After removing the center of mass motion, we need to solve
\begin{equation}
H_{3b}\Psi^{JM\pi}=E\Psi^{JM\pi},
\label{seq3brel}
\end{equation}
with $E$ the three-body energy measured from the three-body breakup threshold. The function $\Psi^{JM\pi}$ is an eigenstate of the three-body Hamiltonian with total angular momentum  $J$, projection on the $z$-axis $M$ and parity $\pi$.  Variational solutions of Eq.\ (\ref{seq3brel}) with $E<0$ are bound states and with $E>0$ are the so-called pseudostates. 

Aiming at solving Eq.\ (\ref{seq3brel}), we make use of the scaled Jacobi coordinates \cite{DDB03}
\begin{gather}
\vis {x}_k=\sqrt{\mu_{ij}}(\vis{r}_j-\vis{r}_i),\notag\\
 \vis {y}_k=\sqrt{\mu_{(ij)k}}\left( \vis {r}_k-\frac{A_i \vis {r}_i+A_j\vis {r}_j}{A_i+A_j}\right),
\label{jcoord}
\end{gather}
with $(i,j,k)$ a cyclic permutation of $\{1,2,3\}$. Thus, we have three sets of scaled Jacobi coordinates.  The dimensionless reduced masses $\mu_{ij}$ and $\mu_{(ij)k}$ are defined as
\begin{equation}
\mu_{ij}=\frac{A_iA_j}{A_i+A_j},\quad \mu_{(ij)k}=\frac{(A_i+A_j)A_k}{A_i+A_j+A_k}.
\label{redmasses}
\end{equation}
Each of the three sets of scaled Jacobi coordinates  in Eq. \ (\ref{jcoord}) defines the hyperspherical coordinates
\begin{equation} 
\rho^2=x_k^2+y_k^2, \label{rho}
\end{equation}
\begin{equation}
\alpha_k=\arctan\left(\frac{y_k}{x_k}\right);\qquad 0\leq \alpha_k \leq \pi/2,\label{HHangle}
\end{equation} 
\begin{equation}\label{hypcoord}
\Omega_{x_k}=(\theta_{x_k},\varphi_{x_k}),\qquad\Omega_{y_k}=(\theta_{y_k},\varphi_{y_k}),
\end{equation}
where $\rho$ and $\alpha_k$ are called the hyperradius and the hyperangle, respectively.

In the hyperspherical formalism, the eigenstate $\Psi^{JM\pi}$ is expanded in hyperspherical harmonics, ${\cal Y}^{JM}_{\gamma K}(\Omega_{5_k})$, as 
\begin{equation}
\Psi^{JM\pi}(\rho,\Omega_{5_k})=\rho^{-5/2}\sum_{K=0}^{\infty}\sum_{\gamma} \chi^{J\pi}_{\gamma K}(\rho) {\cal Y}^{JM}_{\gamma K}(\Omega_{5_k}),
\label{pwf}
\end{equation}
where $\chi^{J\pi}_{\gamma K}(\rho)$ is the hyperradial wave function and the hyperspherical harmonics ${\cal Y}^{JM}_{\gamma K}(\Omega_{5_k})$ are given by
\begin{equation}
\mathcal{Y}^{JM}_{\gamma K}(\Omega_{5_k})=\left[\mathcal{Y}^{l_xl_yK}_{L}(\Omega_{5_k})\otimes\chi_S\right]^{JM},
\end{equation}
with $\Omega_{5_k}=(\Omega_{x_k},\Omega_{y_k},\alpha)$ and $\chi_S$ the spin wave function of the three bodies.  The term ${\cal Y}_{l_xl_yK}^{LM_L}(\Omega_{5})$ is defined as the following coupling
\begin{equation}
{\cal Y}_{l_xl_yK}^{LM_L}(\Omega_{5})=\phi_K^{l_xl_y}(\alpha)\left[Y_{l_x}(\Omega_x)\otimes Y_{l_y}(\Omega_y)\right]^{LM_L},
\label{hharml}
\end{equation}
where we have omitted the subindex $k$ and where the function $\phi_K^{l_xl_y}(\alpha)$ is given in Ref.\    \cite{DDB03}.

Index $\gamma$ in Eq.\ (\ref{pwf}) stands for $\gamma=(l_x,l_y,L,S)$, with $l_x$ and $l_y$ the orbital angular momenta associated with a set of Jacobi coordinates in Eq.\ (\ref{jcoord}), $L$ is the total orbital angular momentum coupled to $l_x$ and $l_y$, and $S$ is the total spin of the clusters. The parity of a state is given by $\pi=(-1)^{K}$. In numerical calculations, the hypermomentum $K$ is truncated at a $K_{\text{max}}$ value.

After inserting Eq.\ (\ref{pwf}) into the Schrödinger equation for two-cluster non-local potentials only, we end up with the set of coupled differential equations \cite{TBD05}
\begin{gather}
\left(-\frac{\hbar^2}{2m_N}
\left[\dfrac{d^2}{d\rho^2}-\dfrac{(K+3/2)(K+5/2)}{\rho^2}\right]-E\right)
\chi^{J\pi}_{\gamma K}(\rho)\notag\\
+\sum_{K' \gamma'}\int_{0}^{\infty}W_{\gamma K,\gamma' K'}(\rho,\rho')\chi^{J\pi}_{\gamma' K'}(\rho')d\rho'=0.
\label{cde}
\end{gather}
The non-local kernel $W_{\gamma K,\gamma' K'}(\rho,\rho')$ in Eq. \ (\ref{cde}) is given by
\begin{align}
W_{\gamma K,\gamma' K'}(\rho,\rho')=\sum_{k=1}^{3}\sum_{l_x''l_y''}&(\rho\rho')^{-3/2}\mu_{ij}^{-3/2}\delta_{LL'}\delta_{SS'}\nonumber\\
&\times\langle k,l_x''l_y''|i,l_xl_y\rangle_{KL}\nonumber\\
&\times\langle k,l_x''l_y''|i,l'_xl'_y\rangle_{KL}\mathcal{I}(\rho,\rho'),\nonumber\\
\label{nlkernel}
\end{align}
with the integral $\mathcal{I}(\rho,\rho')$ defined as
\begin{align}
\mathcal{I}(\rho,\rho')=\int_0^{\text{min}(\rho,\rho')}&W_{k}^{l_x''}\left( \frac{x}{\sqrt{\mu_{ij}}},\frac{x'}{\sqrt{\mu_{ij}}}\right)\nonumber\\
&\phi_K^{l_x''l_y''}(\alpha)\phi_{K'}^{l_x''l_y''}(\alpha')xx'y^2dy.
\label{nlintegral}
\end{align}
In Eq.\ (\ref{nlintegral}), the $W_{k}^{l_x''}\left( \frac{x}{\sqrt{\mu_{ij}}},\frac{x'}{\sqrt{\mu_{ij}}}\right)$ term is a non-local cluster-cluster interaction, $x=(\rho^2-y^2)^{1/2}$,  $x'=(\rho'^2-y^2)^{1/2}$, $\alpha=\arctan(y/x)$, $\alpha'=\arctan(y/x')$. The $\langle k,l_x''l_y''|i,l_xl_y\rangle_{KL}$ in Eq.\ (\ref{nlkernel}) are the Raynal-Revai coefficients \cite{RR70}. These coefficients allow to transform the hyperspherical harmonics that depend on the set of coordinates $i$ to the set of coordinates $k$.
\subsection{Diagonalization of the Hamiltonian with the Lagrange mesh technique}
The Lagrange mesh technique is a variational calculation on a mesh \cite{Ba15}. This efficient technique simplifies the calculations of the Hamiltonian  matrix elements, when they are  computed at the Gauss approximation associated with the mesh. Therefore, with this technique the kinetic matrix elements become analytical and the matrix elements for local potentials are diagonal and evaluated at the mesh points.

Let us use the Lagrange-mesh technique to solve Eq.\ (\ref{cde}). That is, we expand the hyperradial wave function $\chi^{J\pi}_{\gamma K}(\rho)$ over a set of $N$ regularized Lagrange-Laguerre functions $\hat{f}_{n}\left(\frac{\rho}{h}\right)$ as
\begin{equation}
\chi^{J\pi}_{\gamma K}(\rho)=\frac{1}{\sqrt{h}}\sum_{n=1}^N C_{\gamma Kn}^{J\pi}\hat{f}_{n}\left(\frac{\rho}{h}\right),
\label{rexpansion}
\end{equation}
where the $C_{\gamma Kn}^{J\pi}$ are the coefficients of the expansion, $h$ is a scaling parameter that allows to adjust the mesh to the size of  the physical system, and the basis functions $\hat{f}_{n}\left(\frac{\rho}{h}\right)$ are given in Ref.\    \cite{DDB03}.

After inserting the expansion (\ref{rexpansion}) into the set of coupled differential equations (\ref{cde}) and projecting onto  $\frac{1}{\sqrt{h}}\hat{f}_n\left(\frac{\rho}{h}\right)$, we get the eigenvalue problem
\begin{align}
\sum_{\gamma'K'n'}\bigl(&T^{J\pi}_{\gamma Kn,\gamma'K'n'}+W_{\gamma Kn,\gamma' K'n'}\nonumber\\
&\qquad\qquad-E\delta_{\gamma\gamma'}\delta_{KK'}\delta_{nn'}\bigr)C_{\gamma' K'n'}^{J\pi}=0,
\label{nloc-eigen}
\end{align}
with $T^{J\pi}_{\gamma Kn,\gamma'K'n'}$ defined as
\begin{align}
T^{J\pi}_{\gamma Kn,\gamma'K'n'}=\frac{\hbar^2}{2m_N}\biggl(&\frac{\hat{T}^{G}_{nn'}}{h^2}+\nonumber\\
&\frac{(K+ \frac{3}{2})(K+\frac{5}{2})}{u_n^2h^2}\delta_{nn'}\biggr)\delta_{\gamma\gamma'}\delta_{KK'}.
\end{align}
The kinetic matrix elements $\hat{T}^{G}_{nn'}$ of the Lagrange-Laguerre basis are given in Ref.\    \cite{DDB03}. The potential matrix elements $W_{\gamma Kn,\gamma' K'n'}$ become
\begin{align}
&W_{\gamma Kn,\gamma' K'n'}\nonumber\\
&=\frac{1}{h}\int_0^{\infty}d\rho\int_0^{\infty}d\rho'\hat{f}_n\left(\frac{\rho}{h}\right)
W_{\gamma K,\gamma' K'}(\rho,\rho')\hat{f}_{n'}\left(\frac{\rho'}{h}\right),
\end{align}
where the numerical computation is performed following the appendix of Ref.\    \cite{TBD05}.
\section{Two-body EFT potentials} 
\label{sec_2BEFT}
\subsection{Derivation}
The construction of an EFT starts with the finding of the most general Lagrangian of the system. This Lagrangian must be consistent with all the symmetries in terms of the relevant effective degrees of freedom in the energy regime considered.  Galilean-invariant operators in the Lagrangian depend on the relative momenta $\hbar\vis{k}$ and $\hbar\vis{k'}$, only. Thus, the potential can be written as a series of contact interactions and their derivatives, i.e., at low energies in momentum space, we can write 
\begin{equation}
V(\vis{k},\vis{k'})=\sum_{l=0}^{\infty}(2l+1)V_l(k,k')P_l(\hat{\vis{k}},\hat{\vis{k}}'),
\label{pwepot}
\end{equation}
where $P_l$ is a Legendre polynomial of degree $l$ and where the partial wave term of the potential $V_l$ is written as 
\begin{equation}
V_l(k,k')=k^lk'^lg(k)g(k')\sum_{\alpha\beta=0}^{1}k^{2\alpha}\lambda_{\alpha\beta}k'^{2\beta}.
\label{pwavepot}
\end{equation}
The matrix elements $\lambda_{\alpha\beta}$ are defined in the matrix
\begin{equation}
\lambda=
\begin{pmatrix}
\lambda_0 & \lambda_1\\
\lambda_1 & 0 \\
\end{pmatrix}
.
\label{lmatrix}
\end{equation}
Note that the limits of the sums over the indexes $\alpha$ and $\beta$ in Eq.\ (\ref{pwavepot}) could be larger than one. However, we are restricting these sums up to $k^2$ terms as it will be explained later.

The factor
\begin{equation}
g(k)=e^{-(k/\Lambda)^{2m}},
\label{regfun}
\end{equation}
in Eq.\ (\ref{pwavepot}) is a regularization function, where $\Lambda$ is called the cutoff parameter. This parameter suppresses high momenta components, that is, $g(k)\to 0$ when $\Lambda\to\infty$.  However to get $\lambda_0$ and $\lambda_1$ finite, $\Lambda$ must also be finite and limited by the Wigner bound \cite{Wi55}. In principle, observables should not depend on $\Lambda$.  The coefficients $\lambda_0$ and $\lambda_1$ in Eq.\ (\ref{lmatrix}) are the so-called low energy constants (LECs) that capture the effects from high-energy physics. 

The main idea is to find the LECs in Eq.\ (\ref{pwavepot}) from the matching with a low energy theory.  To this end,  we make use of the  partial wave expansion of the Lippmann-Schwinger equation of the T-matrix \cite{Jo75}
\begin{align}
T_l(k,k')=&V_l(k,k')+\frac{1}{2\pi^2}\times\nonumber\\
&\int_0^{\infty}dqq^2V_l(k,q)\dfrac{1}{\frac{\hbar^2k^2}{2\mu}-\frac{\hbar^2q^2}{2\mu}+i\epsilon}T_l(q,k'),
\label{LSTmatrix}
\end{align}
where $\mu$ is the reduced mass of the two clusters.  Here, the $T$-matrix is expanded as in Eq.\ (\ref{pwepot}) and the free state $|\vis{k}\rangle$ is normalized as 
\begin{equation}
\langle\vis{k}|\vis{k'}\rangle=(2\pi)^3\delta(\vis{k}-\vis{k'}).
\end{equation}

The LECs $\lambda_0$ and $\lambda_1$ are found when the potential (\ref{pwavepot}) is introduced in 
Eq.\ (\ref{LSTmatrix}) and it is matched with the on-shell partial wave of the $T$-matrix $T_l$
\begin{equation}
T_l=-\frac{2\pi\hbar^2}{\mu}\left(-\frac{1}{a_l}+\frac{1}{2}r_lk^2-ik^{2l+1}\right)^{-1}k^{2l}.
\label{EREtmatrix}
\end{equation}
Equation (\ref{EREtmatrix}) is derived from the the effective range expansion (ERE) for uncharged cluster-cluster interactions \cite{Be49,BJ49}
\begin{equation}
k^{2l+1}\cot\delta_l=-\frac{1}{a_l}+\frac{1}{2}r_lk^2+\cdots.
\label{EREeq}
\end{equation}
In Eq.\  (\ref{EREeq}) $\delta_l$ is the phase-shift for the elastic scattering in the partial wave $l$. The $a_l$ and $r_l$ are the so-called scattering length and effective range parameters, in units of fm$^{2l+1}$ and fm$^{-2l+1}$, respectively.  We consider terms up to the effective range i.e., up to $k^2$,  which limits the indexes of the sums in Eq.\ (\ref{pwavepot}) up to 1.  Once the experimental values of $a_l$ and $r_l$ are introduced in Eq. \ (\ref{EREtmatrix}) in combination with an appropriate power counting, the LECs $\lambda_0$ and $\lambda_1$ are obtained (see the Appendix for a summary and Refs. \ \cite{FA20,F22,CFJ24} for details).

Equation (\ref{nlintegral})  involves partial waves of cluster-cluster potentials in coordinate representation. They can be obtained from the EFT potentials of the form (\ref{pwepot}) by the double Fourier transform
\begin{equation}
V(\vis{r},\vis{r'})=\frac{1}{(2\pi)^3}\int d\vis{k}d\vis{k'}e^{i(\vis{r}\cdot \vis{k}-\vis{r'}\cdot \vis{k'})}V(\vis{k},\vis{k'}).
\end{equation}
Expanding $V(\vis{r},\vis{r'})$ and $V(\vis{k},\vis{k'})$ in partial waves as shown in Eq.\  (\ref{pwepot}), and with the help of the plane wave expansion in spherical waves,  we have that each partial wave satisfies
\begin{align}
&V_l(r,r')\nonumber\\
&=\frac{2}{\pi}\int_0^{\infty}\int_0^{\infty}dkdk'k^2k'^2j_l(kr)V_l(k,k')j_{l}(k'r').
\label{DFT}
\end{align}
In Eq.\ (\ref{DFT}) $j_l(kr)$ is a spherical Bessel function of the first kind.
\subsection{Forbidden states}
The concept of forbidden states shows up in  many-body theories of fermions, since the full antisymmetrization of the wave function must be taken into account.  For two-cluster nuclei, the Pauli principle can be considered approximately by choosing deep nucleus-nucleus interactions containing unphysical bound states that simulate the forbidden states \cite{BFW77}. These forbidden states are located at energies $E_F$ lower than the physical ones.  

When we are treating three-body nuclei, the two-body forbidden states add spurious eigenvalues associated with the  three-body Hamiltonian that need to be removed.  Different techniques such as the supersymmetry transform of the nucleus-nucleus potential \cite{Ba87} or the projection technique \cite{KP78} have been introduced to remove forbidden states in local nucleus-nucleus potentials.  These transformations keep the phase shifts unaffected.

In this paper,  we extend the projection technique \cite{KP78} to non local potentials by substituting $V_l(k,k')$ in Eq.\  (\ref{pwavepot}), for the $l=0$ $\alpha-n$ potential, by
\begin{equation}
V_l(k,k')=V_l(k,k')+\Gamma\phi_l(k)\phi_l(k'),
\label{pot_proj}
\end{equation}
where $\Gamma$ is a large constant energy value (typically $\Gamma\sim10^3 -10^9$ MeV).  The function $\phi_l(k)$ is a forbidden state or an eigenstate of the nucleus-nucleus Hamiltonian with eigenvalue $E_F$.  Thus,  $\phi_l(k)$ is orthogonal to the physical bound states. 

We should mention that the explicit inclusion of antisymmetrization effects are not a necessity in a cluster EFT approach, since they are determined by wave function overlaps governed by the same EFT expansion of the potential \cite{HJP17,HKK20}. However, eliminating the deep bound state by the choice of a small cutoff ($\Lambda^{0}_{\alpha n} \le$ 100 MeV) results in a much narrower range of agreement with phase shift data ($\le$ 1 MeV) \cite{Yl24}. Here, we have chosen to extend that range to about 5 MeV, at the price of having to eliminate the forbidden state, to have larger cutoff values in the theory.
\section{E1 strength distribution}
\label{sec_E1}
Electric dipole excitations of weakly bound nuclei with just one bound state go directly to the continuum. If such nuclei are made of two-clusters, the calculation of the E1 strength distribution is rather simple \cite{SLY03}. However, for three-body nuclei, the computation of the three-body continuum with the correct asymptotic behavior, which is needed to compute E1 strength distributions, may involve rather heavy calculations such as solving the Faddeev equations in the continuum \cite{GSW05} or the $R$-matrix method \cite{DTB06}.  There are other methods that extend the application of bound state variational calculations to the continuum. They are the complex scaling \cite{AMK06,DPB12}, particularly suitable to treat resonant continuum regions, and integral transform (LIT) methods also in non-resonant energy regions \cite{CS98,Ro20}. Another widely used technique is the so-called pseudostate approach \cite{RAG05,PBD11,DPB12, CSF20}.

In Refs. \ \cite{PBD11,DPB12} the E1 strength distribution of $^6$He is computed in a three-body model with hyperspherical coordinates and two-body local potentials. The complex scaling and the pseudostate methods are tested with a computation that includes the correct asymptotic behavior of the continuum. These references show very good agreement among the computation with the exact continuum and the discretized approaches, although some ambiguity exists in the pseudostate method due to the smoothing technique necessary to derive continuous distributions.  In the following we briefly explain the pseudostate method and clarify that no ambiguity exists when it is used within an integral transform approach. 

We start following the definition of the three-body dipole strength distribution \cite{BCD09}
\begin{align}
\frac{dB_{E1}}{dE}(E)&=\frac{1}{2J_0+1}\delta(E-{\cal E})\nonumber\\
&\sum_{JM\pi M_0\mu}\sum_{\gamma_{\omega}K_{\omega}}
\left|\langle\mathcal{K}\Psi^{JM\pi}_{\gamma_{\omega}K_{\omega}}({\cal E})|\mathcal{M}^{(E1)}_\mu|\Psi^{J_0M_0\pi_0}\rangle\right|^2,
\label{E1exact}
\end{align}
where $\Psi^{J_0M_0\pi_0}$ is the initial bound state characterized by total angular momentum, projection on the $z$ axis and parity,  $J_0$,  $M_0$,  and $\pi_0$. respectively  The final unbound state is represented by $\Psi^{JM\pi}_{\gamma_{\omega}K_{\omega}}(\varepsilon)$,  with total angular momentum $J$,  angular momentum projection $M$ and parity $\pi$, where $\mathcal{K}$ is the time-reversal operator and $E$ is the excitation energy defined from the three-body breakup threshold.  Note that here we have defined the response function with a Dirac Delta function to be further introduced into an integral. 

The electric dipole operator for a $\text{core}+n+n$ system, with the Jacobi coordinate $\vis{x}$ along the $n-n$ motion (see Ref.\ \cite{DDB03} for details), is given by 
\begin{equation}
\mathcal{M}^{E1}_{\mu}(\alpha,\rho)=eZ_c\left(\frac{2}{AA_c}\right)^{1/2}\rho\sin\alpha Y_1^\mu(\Omega_y),
\end{equation}
with $A_c$ and $eZ_c$ the nucleon number and charge of the core, and $A=A_c+2$.

On the other hand,  an observable in an experiment is measured with a resolution defined by the experimental apparatus \cite{Le12}. The latter can be represented in various forms.  We will use a generical function $K_{resol}(E,{\cal E},\Gamma...)$, where $ {\cal E},\Gamma...$ are the parameters of the resolution function. For example, if it is represented by a Gaussian or a Lorentzian function, ${\cal E} $ and $\Gamma   $ could be the center and the width of those functions.  In principle $ \Gamma $ could also depend on $E$.  We take it constant for simplicity of the presentation.

Therefore, because of the experimental resolution, one actually measures the following integral transform of the dipole strength
\begin{equation}
\frac{dB_{E1}}{dE}({\cal E})=\int dE \frac{dB_{E1}}{dE}(E)K_{\text{resol}}(E,{\cal E},\Gamma...).
\label{ConvE1}
\end{equation}
Then, introducing the theoretical E1 strength distribution (\ref{E1exact}) into an integral over $E$, replacing the Delta function by a resolution function, and inserting the completeness relation of states that diagonalizes the three-body Hamiltonian, one can show that for the case of the Borromean nucleus $^{6}$He,  that has $J_0^{\pi}=0^+$ and $J^{\pi}=1^-$,  we obtain \cite{HN05,OT17}
\begin{equation}
\frac{dB_{E1}}{d E}=\sum_{\lambda} 
|\langle\Psi^{1M-}_{\lambda}({\cal E_{\lambda}})|\mathcal{M}^{E1}|\Psi^{00+}\rangle|^2K_{\text{resol}}(E,{\cal E_{\lambda}},\Gamma...).
\label{DiscE1}
\end{equation}
The $\Psi^{1M-}_{\lambda}({\cal E_{\lambda}})$ are pseudostates or eigenstates of the three-body Hamiltonian at positive energies ${\cal E_{\lambda}}$, for the partial wave $J^{\pi}=1^-$.  

What is important here is that for general positive definite Kernels such as Lorentzians or Gaussians, and the easily verified conditions about the existence of the transform and of the total strength, a theorem ensures that the transform can be calculated by means of finite norm functions (the pseudostates) \cite{HN05,OT17}. The smoothed discretized distribution, namely the integral transform, converges exactly to the experimental folded strength distribution, at sufficiently high number of pseudostates. 
It has been noticed that the experimental smoothing procedure excludes higher-frequency oscillations, which therefore are irrelevant.  On the other hand, theoretical results that show isolated Lorentzian/Gaussian peaks may indicate that the corresponding structure is not sufficiently resolved.  Thus, an enlargement of the basis space is necessary.

Response functions can be obtained from the inversion of integral transforms \cite{ELO98}. However, we would like to emphasize that if a smooth theoretical result is obtained with the experimental resolution, an inversion of the transform is not necessary. Problems can only arise if the experimental resolution is so small that such a smooth result is not obtained and isolated peaks due to single pseudostates appear, even if the basis is enlarged. In such a case it is more reliable to use an inversion for the transform \cite{ELS19}.
\section{$^{6}$He structure}
\label{sec_6he}
\subsection{Two-body potentials}
The derivation of the LECs $\lambda_0$ and $\lambda_1$ for the two-cluster potential (\ref{pwavepot}) is performed in Refs.  \cite{FA20,F22,CFJ24} and is summarized in the Appendix.  We use the experimental values of the scattering length $a_l$ and the effective range $r_l$ cited in Table \ref{tab_EFP}.  They are taken from Refs.  \cite{CC08,Ma22} for the $n-n$ system, and from Ref.\  \cite{ALR73} for the $\alpha-n$ system.

\begin{table}[htb]
\caption{Experimental scattering length and effective range parameters to obtain two-body EFT potentials.}
\label{tab_EFP}
\begin{ruledtabular}
\begin{tabular}{ l c c c}
System       & $l_j$          & $a_l$ (fm$^{2l+1}$)  & $r_l$  (fm$^{-2l+1}$)\\  
\hline
$n+n$         & $S_0$       &  $-18.630\pm 0.480$   & $\quad 2.870\pm 0.100$ \\
$\alpha+n$ & $S_{1/2}$	&  $\quad\; 2.464\pm 0.004$  & $\quad 1.385\pm 0.041$\\
$\alpha+n$ & $P_{3/2}$	&  $-62.951\pm 0.003$  & $\;-0.882\pm 0.001$\\
\end{tabular}
\end{ruledtabular}
\end{table}

Once the EFT potentials are calculated in momentum space, they are transformed to coordinate representation by performing a double Fourier transform as indicated in Eq.\ (\ref{DFT}).  The integrals are computed numerically with the help of a double Gauss-Legendre quadrature of $N_k=300$ points.  

\begin{figure}[h!]
\begin{center}
\includegraphics[scale=0.73]{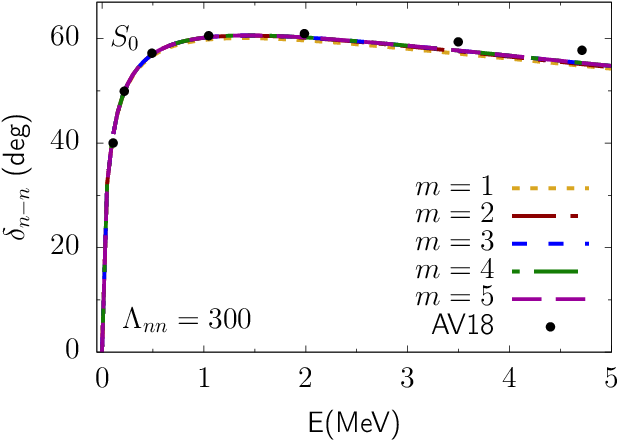}
\includegraphics[scale=0.73]{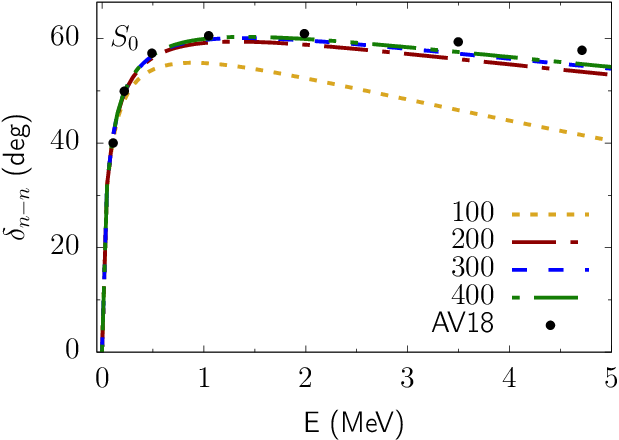}
\end{center}
\caption{Dependence of the EFT $n-n$ phase shift on the exponent of the regularization function (top) and on the cutoff parameter $\Lambda_{nn}$, in MeV, in natural units (bottom). The points are the calculations with the AV18 potential \cite{WSS95}.  }
\label{delta_nn}
\end{figure}

Figure \ref{delta_nn} shows the low energy behavior of the $n-n$ $^{1}S_0$ phase shift computed with the two-body EFT potentials in comparison with a calculation that uses the AV18 potential \cite{WSS95}.  We show the dependence on the exponent $m$ of the Gaussian regularization function (\ref{regfun}) and on the cutoff parameter $\Lambda_{nn}$.  As mentioned before, this parameter should be as large as possible,  limited by the Wigner bound and the physics, in principle, should not depend on it.  We observe a slight dependence on the cutoff parameter apart from $\Lambda_{nn}=100$ MeV.  The low energy phase-shifts for the $S$-wave and $P_{3/2}$ $\alpha-n$ systems are studied in Refs. \ \cite{FA20,F22}.

In coordinate representation, we calculate the phase shifts with the $R$-matrix method for non-local potentials \cite{Ba15}.  
We also check that the Fourier transformed EFT potentials reproduce the experimental scattering lengths $a_l$ and the effective range $r_l$ values.  To this end, we extended Ref.\  \cite{RS13} to compute effective range parameters with the $R$-matrix method for non-local potentials. 

Table \ref{tab_EFTpar} gives the set of EFT potential parameters that are used to compute ground state properties of $^{6}$He, unless mentioned otherwise.  We choose $\Lambda$ values as large as possible avoiding being close to Wigner bounds. 

\begin{table}[htb]
\caption{EFT two-body potential parameters in natural units \footnote{In. S.I units $\lambda_0\to\lambda_0 \hbar c$, $\lambda_1\to\lambda_1 \hbar c$ and $\Lambda\to\Lambda/ \hbar c$, where $\hbar c=197.3$ MeVfm. }.}
\label{tab_EFTpar}
\begin{ruledtabular}
\begin{tabular}{cccccc}
System & $l_j$    & $m$  &  $\Lambda$ (MeV) &$\lambda_0$ (fm$^{2(l+1)}$) & $\lambda_1$ (fm$^{4+2l}$) \\  
\hline
 $n+n$  & $S_0$ & $1$   & $300$ & -5.320  & 1.660\\
$\alpha+n$  & $S_{1/2}$ & $2$   & $400$ & -6.764 & 1.311\\
$\alpha+n$  & $P_{3/2}$ & $1$   & $300$ &  -10.177 &  2.714\\
\end{tabular}
\end{ruledtabular}
\end{table}

On the other hand, the $S$-wave $\alpha-n$ EFT potential binds a forbidden state for $\Lambda^{0}_{\alpha n}>100$ MeV. We remove this state by adding a projector to the potential in momentum space as indicated in Eq.\ (\ref{pot_proj}). The functions $\phi(k)$ are obtaining by solving the Schrödinger equation in momentum space following Ref.\ \cite{LSB12} with $N_p=30$ Lagrange-Laguerre basis functions and scaling parameter $h_p=0.3$ fm.  We have checked that the phase shifts remain unaffected with the addition of the projector potential.  Once we have the potential (\ref{pot_proj}), we proceed to perform its double Fourier transform.  Table \ref{FS_energies} shows the $S$-wave $\alpha-n$ potential parameters for $m=2$ and different cutoff parameters $\Lambda^{0}_{\alpha n}$. We also show the associated forbidden state energies.  These energies agree with the value of 12.38 MeV given by the Kanada et al. local potential \cite{KKN79} used in Ref.\ \cite{DDB03}. 
\begin{table}[htb]
\caption{EFT potential parameters, in natural units, and forbidden state energies for $S$-wave $\alpha-n$ potentials with $m=2$.}
\label{FS_energies}
\begin{ruledtabular}
\begin{tabular}{cccc}
$\Lambda^{0}_{\alpha n}$ (MeV)   & $\lambda_0$  (fm$^{2(l+1)}$) &  $\lambda_1$ (fm$^{2l+4}$) & $E_F$ (MeV) \\  
\hline
$200$ &  $-2.031$  & $-7.331$ & $-5.61$   \\
$300$ & $-6.963$ & $0.807$ & $-12.23$\\
$400$ & $-6.764$ & $1.311$ & $-13.34$\\
$500$ & $-6.290$   & $2.005$  &$-13.84$ \\
\end{tabular}
\end{ruledtabular}
\end{table}

\subsection{$\alpha+n+n$ ground state properties}\label{gs}
The two-body potentials with parameters given in Table \ref{tab_EFTpar} are used in Eq.\ (\ref{nloc-eigen}) to compute the ground state energy, $E_0$, of $^6$He.  We solve the diagonalization problem as shown in Ref.\    \cite{TBD05} with $N=60$ mesh points and scaling parameter $h=0.3$ fm. The integrals over $x$ and $x'$ are computed with a  Gauss-Laguerre quadrature with $N_2=60$ and $h_2=0.05$ fm. The integral over $y$ is performed with $N_y=500$ points and a step of $h_y=0.02$ fm.

The two-body potentials alone do no bind the system. Therefore, we introduce the three-body force 
\begin{equation}
V_3(\rho)=-V_{03}e^{-\left(\frac{\rho}{\rho_0}\right)^2},
\label{3Binteraction}
\end{equation}
which corresponds to a $S$-wave LO cluster EFT potential in coordinate representation.  In Eq.\ (\ref{3Binteraction}) $\rho_0$ makes the role of a three-body cutoff parameter. The introduction of this force allows to reproduce the correct experimental ground state energy and to remove dependencies on this energy of the cutoff parameter of the two-body EFT potentials. 

\begin{figure}[h!]
\begin{center}
\includegraphics[scale=0.73]{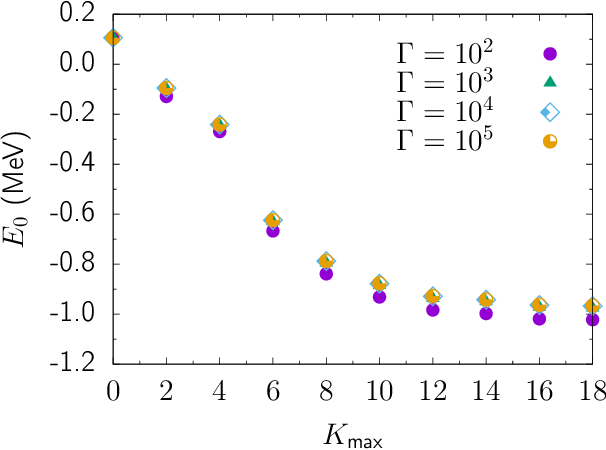}
\end{center}
\caption{Convergence with $K_{\text{max}}$ of the ground state energy of $^6$He for different values of the projector parameter $\Gamma$ in MeV.}
\label{Gamma-dep}
\end{figure}

In Fig.  \ref{Gamma-dep}, we study the convergence of $E_0$  with respect to the maximum $K$ value,  $K_{\text{max}}$, in the expansion (\ref{pwf}) of the three-body wave function.  We also check the dependence of this energy on the projector $\Gamma$ at eliminating the forbidden state in the $S_{1/2}$ non-local $\alpha-n$ potential. As it is the case for local potentials \cite{DDB03},  $\Gamma$ values at least of 1000 MeV are needed to eliminate dependencies of the ground state energy.  

We also compute the rms radius of the $\alpha+n+n$ system through the expression \cite{ZDF93}
\begin{equation}
\langle r^2 \rangle_{^6\text{He}}=\frac{1}{6}\langle\Psi^{J_0M_0\pi_0}|\rho^2|\Psi^{J_0M_0\pi_0}\rangle +\frac{2}{3}\langle r^2 \rangle_{\alpha},
\end{equation}
with $\Psi^{J_0M_0\pi_0}$ the ground state of $^6$He and $\sqrt{\langle r^2 \rangle}_{\alpha}$ the rms radius of the $^4$He nucleus.  For this radius we take the experimental value of $1.463(6)$ fm given in Ref.\ \cite{Si08}.  

In Table \ref{tab_GS} we show the sensitivity of the computed rms radius with the choice of the three-body potential.  The percentage difference between the predicted values is up to $30\%$.  The values of $\rho_0$ that provide radii closer to the experimental values are $\rho_0=4$ fm, $\rho_0=4.5$ fm and $\rho_0=5$ fm, with percentage differences of $2.5\%$, $0.24\%$ and $3.4\%$, respectively.

Because of the definition of $\rho$, Eqs. ~(\ref{jcoord})--(\ref{rho}), the hyperradial cutoff, in line with the cutoffs $\Lambda$ for the two-body potentials, is about 1 fm.  Nonetheless, our smallest value of $\rho_0$ is only 1.5 fm, since for even smaller values of $\rho_0$ the HH convergence becomes very poor. However, extrapolating the results for the rms radius in Table~\ref{tab_GS},  we find that a $\rho_0$ of 1 fm would lead to a rather unrealistic radius.

\begin{table}[htb]
\caption{Ground state energies and rms radii of $^6$He for different three-body forces. The experimental values are $E_0=-0.97546(23)$ MeV \cite{BBC12} and $r_{rms}=2.49(4)$ fm \cite{SNK21} .}
\label{tab_GS}
\begin{ruledtabular}
\begin{tabular}{ c c c c}
  $\rho_0$ (fm)   &  $V_{03}$ (MeV)      & $E_0$ (MeV)  & $r_{rms}$ (fm)\\  
\hline
    $1.5$   & $-518. 2$    & $-0.985$    & $1.790$ \\
    $2.0$   &  $-200.8$   &  $-0.973$  & $2.137$  \\
    $2.5$   &  $-87.5$   &  $-0.976$  & $2.215$  \\ 
    $3.0$   &  $-49.5$   &  $-0.981$  & $2.286$  \\   
    $3.5$   &  $-32.9$   &  $-0.975$  & $2.358$  \\   
    $4.0$   &  $-24.3$   &  $-0.983$  & $2.428$  \\  
    $4.5$   &  $-19.3$   &  $-0.981$  & $2.500$  \\     
    $5.0$   &  $-15.9$   &  $-0.976$  & $2.574$  \\  
\end{tabular}
\end{ruledtabular}
\end{table}

\subsection{Electric dipole strength distribution}
In this section we compute the electric dipole strength distribution of $^6$He using Eq.\ (\ref{DiscE1}).  We employ a Gaussian smoothing distribution with $\sigma=\left(0.027+0.177E^{0.654}\right)$ MeV, which corresponds to the experimental energy resolution \cite{SNK21}.  This distribution is normalized to unity from $0$ to $\infty$. We use the same Hamiltonian to compute the ground and the $1^-$ discrete states.  

\begin{figure}[h!]
\begin{center}
\includegraphics[scale=0.73]{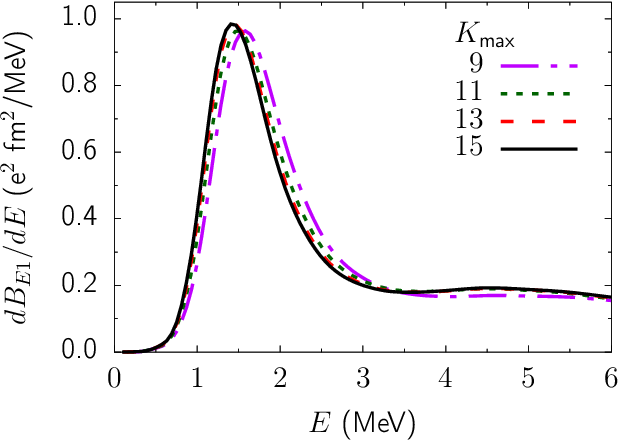}
\end{center}
\caption{Convergence of the E1 strength distribution of $^6$He with $K_{\text{max}}$. }
\label{Kconv}
\end{figure}

\begin{figure}[h!]
\begin{center}
\includegraphics[scale=0.73]{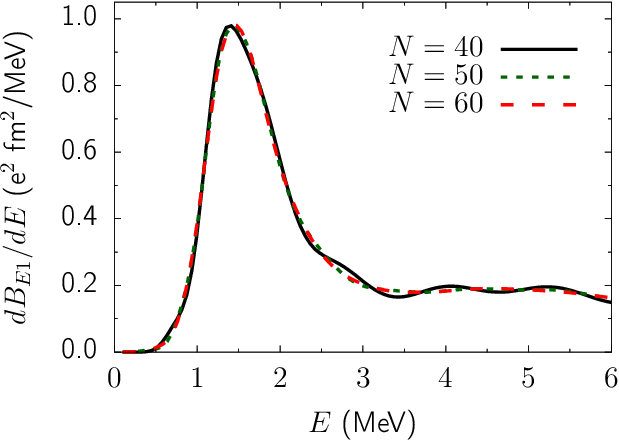}
\end{center}
\caption{Convergence of the E1 strength distribution of $^6$He with the number of the Lagrange-Laguerre basis functions $N$.}
\label{Nconv}
\end{figure}

\begin{figure}[h!]
\begin{center}
\includegraphics[scale=0.73]{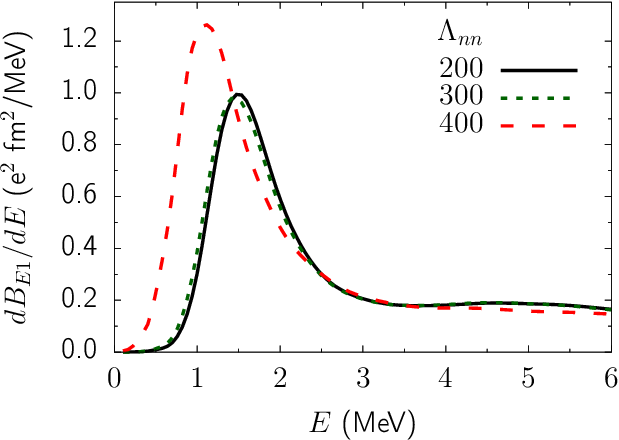}
\includegraphics[scale=0.73]{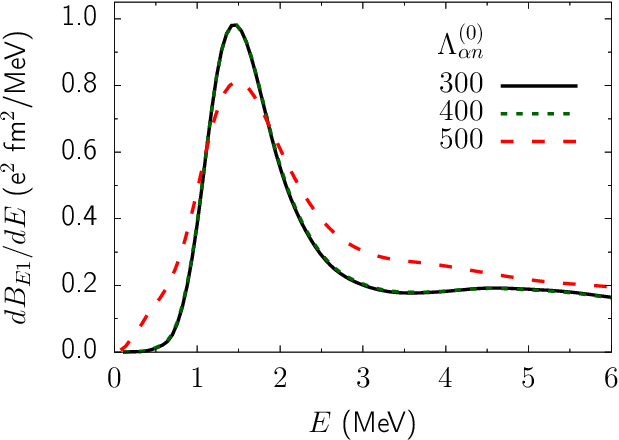}
\includegraphics[scale=0.73]{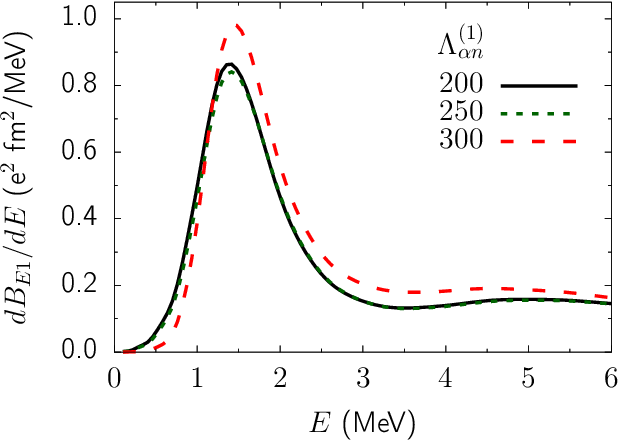}
\end{center}
\caption{Influence of the cutoff parameter, labels in MeV, in natural units, of the two-body potentials, on the E1 strength distribution of $^6$He.}
\label{E1-lambda}
\end{figure}

\begin{figure}[h!]
\begin{center}
\includegraphics[scale=0.78]{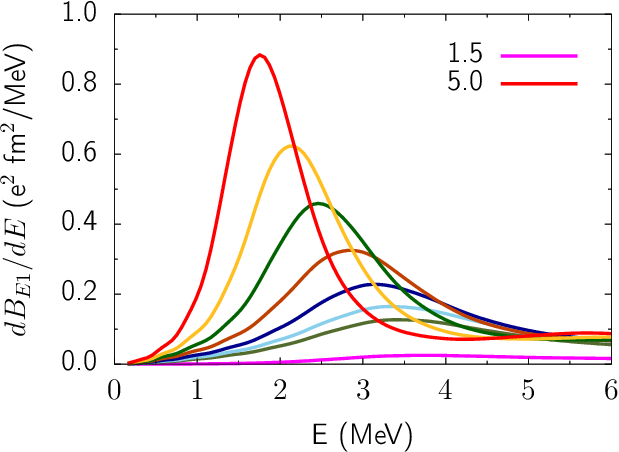}
\end{center}
\caption{Dependence of the E1 strength distribution of $^6$He on the cutoff parameter $\rho_0$ of the three-body force. The values of $\rho_{0}$ (in fm) vary from 1.5  to 5  in 0.5 units. The curves are shown in ascending order in $\rho_{0}$ from left to right. }
\label{E1-rho}
\end{figure}

In Figs. \ref{Kconv}, \ref{Nconv} and \ref{E1-lambda}, we show the convergence of the E1 strength distribution with the maximum value of the hypermomentum $K_{\text{max}}$ of the $1^-$ pseudostates, the number $N$ of the Lagrange-Laguerre basis functions involved in the diagonalization of the three-body Hamiltonian, and the cutoff parameters of the two-body potentials.  We choose a three-body potential with $\rho_0=5$ fm.  For the sake of reducing computational times, we compute the integrals over $y$ with $N_y=250$ points and a step of $h_y=0.05$ fm. The integrals over $x$ and $x'$ are performed with $N_2=30$ and $h_2=0.3$ fm.  Such choices provide an E1 strength distribution shifted to lower energies in comparison with a full converged calculation as shown in Fig.  \ref{E1-rho}.

In Fig. \ref{Kconv} we observe that we achieve convergence at $K_{\text{max}}=13$. In Fig.  \ref{Nconv} the number of pseudostates involved corresponds to 242,  395 and 525 for $N=40$,  $N=50$ and $N=60$, respectively. As isolated resonances are not obtained at increasing the number of pseudostates in the sum (\ref{DiscE1}), we do not need to perform an inversion of the integral transform (\ref{ConvE1}).

\begin{figure}[h!]
\begin{center}
\includegraphics[scale=0.58]{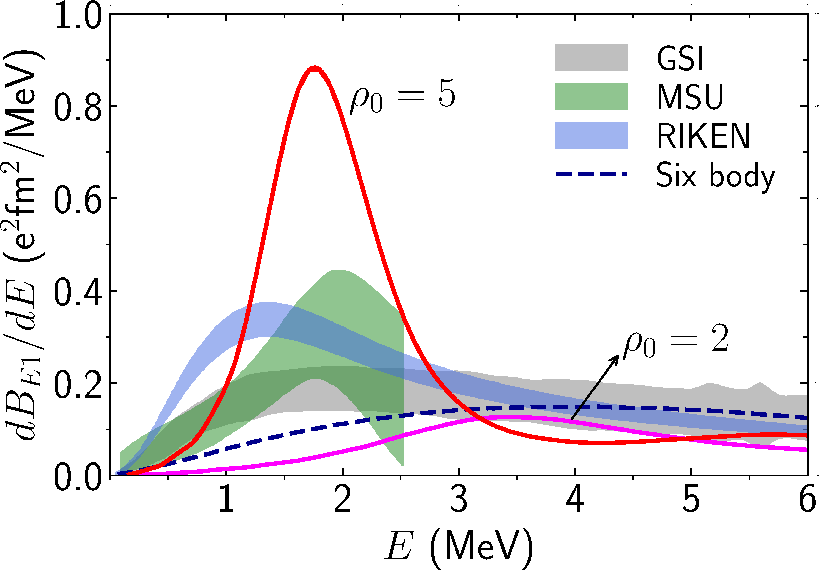}
\end{center}
\caption{Present work theoretical calculations of the E1 strength distribution of $^6$He for $\rho_{03}=2$ and $\rho_{03}=5$, in fm,  (solid lines), in comparison with the available experimental data (shade areas labeled GSI, MSU and RIKEN taken from Refs. \ \cite{AAA99,WGK02,SNK21}). The dashed line is the six-body theoretical calculation of Ref. \ \cite{BB04}. }
\label{E1-sameH}
\end{figure}

Aiming at assessing the effect of the cutoff parameter of the two-cluster potentials on the E1 strength distribution, we show this distribution in Fig.  \ref{E1-lambda} for different values of the $\Lambda_{nn}$,  $\Lambda^{0}_{\alpha n}$  and $\Lambda^{1}_{\alpha n}$ cutoff parameters of the $S_0$ $n-n$,  $S_{1/2}$ $\alpha-n$ and $P_{3/2}$ $\alpha-n$ potentials, respectively.  We observe a strong dependence on the E1 transition distribution for the values $\Lambda_{nn}=400$ MeV,  $\Lambda^{0}_{\alpha n}=500$ MeV and $\Lambda^{1}_{\alpha n}=300$ MeV,  which are close to the Wigner bounds 410 MeV, 510 MeV and 340 MeV of the $S_0$ $n-n$,  $S_{1/2}$ $\alpha-n$ and $P_{3/2}$ $\alpha-n$ potentials, respectively. We also find that the ground state of  $^6$He exhibits a slow convergence with $K_{\text{max}}$ ($K_{\text{max}}=30$), when the two-body potentials have a cutoff parameter that tends to a Wigner bound.  In addition, we check the dependence of the E1 strength with the exponent $m$ of the regularization function (\ref{regfun}). We get that this quantity differs at most $10\%$ for all potentials in the peak region. In general, the EFT $^6$He E1 strength distributions are in good agreement with the computation of the three-body model with phenomenological two-body local potentials of Ref.\ \cite{BCD09}.

\begin{figure}[h!]
\begin{center}
\includegraphics[scale=0.73]{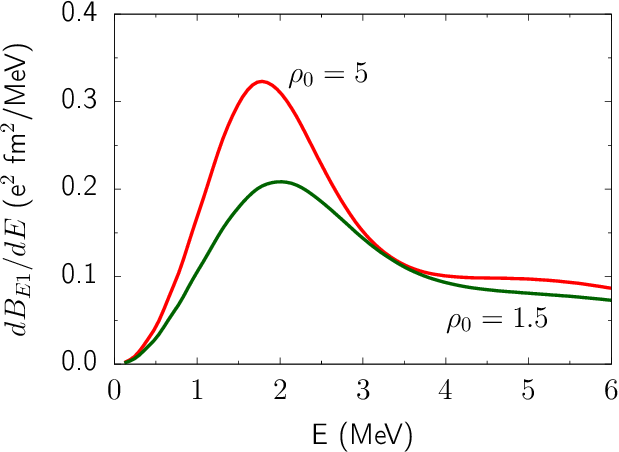}
\end{center}
\caption{Sensitivity of the E1 strength distribution of $^6$He with $\rho_0$ (in fm), when two-body local phenomenological potentials are used. }
\label{E1-local}
\end{figure}

In Fig.  \ref{E1-rho}, we show the dependence of the electric dipole strength distribution with $\rho_0$, i.e., with the cutoff parameter of the $S$-wave LO three-body potential.  The two-body potentials parameters are given in Table \ref{tab_EFTpar} and the numerical conditions are described in Sec.  \ref{gs}.  We vary $\rho_0$ from 1.5 fm to 5 fm in 0.5 units.  Our calculations show a strong dependence on the three-body cutoff parameter.  It is interesting that a small $\rho_0$ of 1.5 fm, which is more in line with the two-body cutoffs, leads to a low and very flat strength distribution. We briefly investigated also the state dependence of the three-body force by choosing a different strength $V_{03}$ for the three-body force in the $1^-$ channel such that the peak of the E1 strength distribution is shifted to the peak position of the recent Riken experiment. It leads to somewhat reduced dependence of the three-body force on the peak height, but the peak widths turn out to be much higher than found in experiment. This behavior indicates that inclusion of higher-order terms in the effective field theory is required to eliminate the $\rho_0$ dependence.

In Fig. \ref{E1-sameH} we compare the available experimental data of the $^6$He E1 strength distribution with the computed distributions for $\rho_0=2$ fm and $\rho_0=5$ fm. They correspond to ground state wave functions that provide a percentage difference with the experimental value of the rms radius of $14.2\%$ and $3.4\%$, respectively.  The computation with $\rho_0=2$ fm  approaches the six-body computation of Ref. \ \cite{BB04}, while the behavior of the curve with $\rho_0=5$ fm agrees roughly with the experimental data of  Ref. \ \cite{WGK02,SNK21}. Although, the strength is much higher. 

It is well possible that an inclusion of higher order terms will reduce the importance of the three-body force. In this context
it is interesting to compare with a phenomenological ansatz for the two-body potentials which includes also higher two-body partial waves. 
Let us recall that in the LO calculation only interactions in $nn$ $s$-wave and $\alpha n$ $s$- and $p$-wave are present.  Other partial waves will appear by the inclusion of higher EFT terms.  Therefore, we show in Fig.  \ref{E1-local} the dependence on $\rho_0$ of the E1 strength distribution of $^6$He, when using the three-body interaction (\ref{3Binteraction}).  We use the three-body model of Refs. \ \cite{PBD11, DPB12} with two-body local phenomenological potentials and a supersymmetric transformation to eliminate the forbidden state.  However, for the sake of simplicity in the comparison, we just use $K_{\text{max}}=9$ for the final state.   Higher values will shift the peak position to lower energies.  We eliminate isolated unphysical resonances by choosing a Lagrange-Laguerre mesh with $N=70$.  The depth of the three-body potential is tuned to  $-94$ MeV and to $-2.5$ MeV to reproduce the experimental ground state energy of $^6$He. These values correspond to $\rho_0=1.5$ fm and for $\rho_0=5$ fm, respectively.  The corresponding rms radii are 2.37 fm and 2.47 fm.  Without a three-body force the ground state energy is $E_0=-0.38$ MeV.  This means that the addition of a three-body force only provides a fine tuning 
for the the ground-state energy and its influence on the E1 strength is considerably reduced as shown in Fig.  \ref{E1-local}.
In addition it is interesting to note that  different from the cases in Fig.~\ref{E1-sameH} the peak position almost does not vary
with the three-body cutoff $\rho_0$.

\section{Summary and conclusions}
\label{sec_conclu}
 We have studied the nucleus of $^6$He within a cluster EFT approach. We have taken the $\alpha n$ and $nn$ potentials,
naturally derived in momentum space, and Fourier transformed them to coordinate space. Besides avoiding  complications with the Coulomb interaction, this will allow to introduce the three-body wave functions into existing low-energy reaction codes. This is necessary to achieve  a more fundamental description of reaction processes and to study non-local effects in reaction theory.  

To this end, we have started with the $^6$He nucleus by using a $S_0$ $n-n$, and a $S_{1/2}$ $\alpha-n$ and $P_{3/2}$ $\alpha-n$ EFT potentials. We have used a three-body model in hyperspherical coordinates with the Lagrange-mesh technique.  The model incorporates the two-body non-local interactions that result from the Fourier transformed cluster EFT potentials in momentum space.  We have also addressed the removal of forbidden states in cluster EFT potentials by extending the projection technique \cite{KP78} to the non-local $S_{1/2}$ $\alpha-n$ potential. A LO $S$-wave three-body interaction is introduced in the Hamiltonian to properly bind the $^6$He nucleus and to eliminate dependencies on the two-body cutoff parameter. 

Besides the ground state energy of $^6$He, we have computed its rms radius and the electric dipole strength distribution. For the latter, we have used the pseudostate  approach, which corresponds, under certain conditions, to the integral transform method. In fact, we have compared the results to the experimental data employing the same resolution as in experiment.  Such a controlled   pseudostate approach, which uses the discretization of the continuum, does not represent an approximation, under the conditions of the present calculation, and the obtained results can be compared directly to the data.  The condition, that an increase in the number of pseudostates exhibits a convergent smooth result without isolated resonances, is verified by our calculation. Therefore an inversion of the transform has not even been necessary.

An EFT should not, in principle, depend on the cutoff parameter of the renormalization function. Thus, we have checked this dependence for the two-body potentials as well as for the three-body interaction. We have observed that the E1 strength distribution depends strongly on the two-body cutoff parameter when it is close to the Wigner bound,  requiring a certain care  at choosing this value.  

Our analysis reveals that the  root-mean-square radius of $^6$He,  which characterizes the average properties of the ground state, exhibits a clear dependence on the three-body cutoff.  The theoretical predictions deviate from the experimental data by as much as  $30\%$ in the worst case. The sensitivity with the three-body cutoff is even more pronounced in the E1 strength distribution, since that observable probes both the bound ground state and the $1^{-}$ continuum pseudostates. We attribute the three-body cutoff dependence to missing higher order terms in the EFT potentials.  However, the present model provides a robust starting point for the investigation of Borromean nuclei with more fundamental potentials, in coordinate representation.

On the other hand the experimental situation is also not yet fully settled. The variation of different experimental results is relatively large. In addition the experimental determination of the dipole strength is strongly model dependent  as was pointed out in Ref.\ \cite{PDB12}.

All this makes $^6$He still a very interesting and challenging ground for continuing both theoretical and experimental investigations.
\section*{Acknowledgments}
We thank C. Ji, E. Filandri and D. Baron for helping discussions on the EFTs potentials.  E.  C.  Pinilla gratefully acknowledges the support given  by the INFN.  W.L. acknowledges the financial support of
the European Union - Next Generation EU, Mission
4 Component 1, CUP I53D23001060006, for the PRIN
2022 project “Exploiting separation of scales in nuclear structure and dynamics”. 
\appendix*
\section{LECs under Wigner bound for neutral two-body interactions}
\label{App_LECS}
In this section, we summarize the derivation of the LECs $\lambda_0$ and $\lambda_1$ of the EFT potential (\ref{pwavepot}) in natural units.  We refer the reader to Ref.\ \cite {FA20,F22,CFJ24} for details.

Let us make use of the partial wave expansion of the two-body Lippmann Schwinger equation (\ref{LSTmatrix}), the partial wave potential (\ref{pwavepot}) and a similar expansion for the $T$-matrix 
\begin{equation}
T_l(k, k')=k^lk'^lg(k)g(k')\sum_{i j=0}^1 k^{2 i} \tau_{i j}(E) k^{\prime 2 j}.
\label{tmatrix}
\end{equation}
After introducing Eq.\ (\ref{tmatrix}) and Eq.\ (\ref{pwavepot}) in the Lippmann Schwinger equation (\ref{LSTmatrix}), we have
\begin{align}
\tau_{i j}=&\lambda_{i j}\notag\\
&+\frac{1}{2 \pi^2}\sum_{i' j'=0}^1\lambda_{i i'} \int_0^{\infty}dk k^2\frac{k^{2l+2i'+2j'}}{E-k^2 /2\mu+i \epsilon} g^2(k)\tau_{j' j},
\label{TMindex}
\end{align}
where $g(k)$ is given by Eq. \ (\ref{regfun}). 

Equation (\ref{TMindex}) corresponds to the matrix equation
\begin{equation}
\tau=\lambda+\lambda\Phi^{(l)}\tau,
\label{TMME}
\end{equation}
with the matrix $\lambda$ given by Eq.\ (\ref{lmatrix}) and the matrix $\Phi^{(l)}$ defined as
\begin{equation}
\Phi^{(l)}=
\begin{pmatrix}
\phi^{(l)}_0 &\phi^{(l)}_2\\
\phi^{(l)}_2& \phi^{(l)}_4 \\
\end{pmatrix}
.
\label{phimatrix}
\end{equation}
The matrix elements of $\Phi^{(l)}$ are given by the expression
\begin{equation}
\Phi^{(l)}_{2\nu}=\frac{1}{2 \pi^2}\int_0^{\infty}dk k^2\frac{k^{2l+2\nu}}{E-k^2 /2\mu+i \epsilon} g^2(k),
\end{equation}
with $\nu=0,1,2$. These matrix elements can be written recursively as
\begin{align}
& \phi^{(l)}_{4}=I_{2 l+5}+k^2 I_{2 l+3}+k^4 \phi_0^{(l)}, \\
& \phi^{(l)}_{2}=I_{2 l+3}+k^2 \phi_0^{(l)}, \\
& \phi^{(l)}_{2\nu}=I_{2 l+1}+\cdots+k^{2 l+2} \phi_{-1}^{(l)},
\end{align}
where the integral $I_{n}$ is given by
\begin{equation}
I_{n}=-\frac{\mu}{\pi^2} \int_0^{\infty} d q q^{n-1} g^2(q).
\end{equation}
When the regulator function is defines as in Eq.\ (\ref{regfun}), we have
\begin{equation}
 I_n =-\frac{\mu}{\pi^2}\frac{\Lambda^n}{n}f_{n,m},
\end{equation}
with
\begin{equation}
f_{n, m}=\left(\frac{1}{2}\right)^{\frac{n}{2 m}} \Gamma\left(\frac{n}{2 m}+1\right).
\end{equation}
After solving Eq.\ (\ref{TMME}), rescaling the the LECs to
\begin{equation}
\lambda_0=-\frac{\pi^2}{\mu} \frac{c_0}{\Lambda^{2 l+1}} \quad \lambda_1=-\frac{\pi^2}{\mu} \frac{c_1}{\Lambda^{2 l+3}},
\end{equation}
expanding in powers of $(k/\Lambda)^2$ and comparing with the effective range expression for the on shell $T$-matrix  (\ref{EREtmatrix}), we have expressions for the coefficients $C_0$ and $C_1$ given by
\begin{gather}
c_0=\Biggl[\dfrac{(\frac{f_{3,m}}{3}-\frac{1}{c_1})^2}{f_{1,m}-\frac{\pi}{2a_0\Lambda}}-\frac{f_{5,m}}{5}\Biggr]c_1^2,\nonumber\\
c_1=\Biggr\{1\pm\Biggl[1-\dfrac{(f_{-1,m}-\frac{r_0\pi\Lambda}{4})\frac{f_{3,m}}{3}}{\bigl(f_{1,m}-\frac{\pi}{2a_0\Lambda}\bigr)^2}\Biggr]^{-1/2}\Biggl\}\dfrac{3}{f_{3,m}},
\end{gather}
for the $l=0$ potentials and
\begin{gather}
c_0=\Biggl[\dfrac{(\frac{f_{5,m}}{5}-\frac{1}{c_1})^2}{\frac{f_{3,m}}{3}-\frac{\pi}{2a_1\Lambda^3}}-\frac{f_{7,m}}{7}\Biggr]c_1^2,\nonumber\\
c_1=\Biggr\{1\pm\Biggl[1-\dfrac{(f_{1,m}+\frac{r_1\pi}{4\Lambda})\frac{f_{5,m}}{5}}{\bigl(\frac{f_{3,m}}{3}-\frac{\pi}{2a_1\Lambda^3}\bigr)^2}\Biggr]^{-1/2}\Biggl\}\dfrac{5}{f_{5,m}},
\end{gather}
for the $l=1$ potentials. The $a_l$ and $r_l$ are the experimental scattering length and the effective range associated to the partial wave $l=0,1$.  For each $l$, the $c_1$ constant has two values. The negative root is chosen since it  provides the smaller values of most natural size.

\bibliography{biblio2}

\begin{thebibliography}{57}%
\makeatletter
\providecommand \@ifxundefined [1]{%
 \@ifx{#1\undefined}
}%
\providecommand \@ifnum [1]{%
 \ifnum #1\expandafter \@firstoftwo
 \else \expandafter \@secondoftwo
 \fi
}%
\providecommand \@ifx [1]{%
 \ifx #1\expandafter \@firstoftwo
 \else \expandafter \@secondoftwo
 \fi
}%
\providecommand \natexlab [1]{#1}%
\providecommand \enquote  [1]{``#1''}%
\providecommand \bibnamefont  [1]{#1}%
\providecommand \bibfnamefont [1]{#1}%
\providecommand \citenamefont [1]{#1}%
\providecommand \href@noop [0]{\@secondoftwo}%
\providecommand \href [0]{\begingroup \@sanitize@url \@href}%
\providecommand \@href[1]{\@@startlink{#1}\@@href}%
\providecommand \@@href[1]{\endgroup#1\@@endlink}%
\providecommand \@sanitize@url [0]{\catcode `\\12\catcode `\$12\catcode
  `\&12\catcode `\#12\catcode `\^12\catcode `\_12\catcode `\%12\relax}%
\providecommand \@@startlink[1]{}%
\providecommand \@@endlink[0]{}%
\providecommand \url  [0]{\begingroup\@sanitize@url \@url }%
\providecommand \@url [1]{\endgroup\@href {#1}{\urlprefix }}%
\providecommand \urlprefix  [0]{URL }%
\providecommand \Eprint [0]{\href }%
\providecommand \doibase [0]{https://doi.org/}%
\providecommand \selectlanguage [0]{\@gobble}%
\providecommand \bibinfo  [0]{\@secondoftwo}%
\providecommand \bibfield  [0]{\@secondoftwo}%
\providecommand \translation [1]{[#1]}%
\providecommand \BibitemOpen [0]{}%
\providecommand \bibitemStop [0]{}%
\providecommand \bibitemNoStop [0]{.\EOS\space}%
\providecommand \EOS [0]{\spacefactor3000\relax}%
\providecommand \BibitemShut  [1]{\csname bibitem#1\endcsname}%
\let\auto@bib@innerbib\@empty
\bibitem [{\citenamefont {Weinberg}(1979)}]{We79}%
  \BibitemOpen
  \bibfield  {author} {\bibinfo {author} {\bibfnamefont {S.}~\bibnamefont
  {Weinberg}},\ }\href@noop {} {\bibfield  {journal} {\bibinfo  {journal}
  {Physica A}\ }\textbf {\bibinfo {volume} {96}},\ \bibinfo {pages} {327}
  (\bibinfo {year} {1979})}\BibitemShut {NoStop}%
\bibitem [{\citenamefont {Weinberg}(1980)}]{We80}%
  \BibitemOpen
  \bibfield  {author} {\bibinfo {author} {\bibfnamefont {S.}~\bibnamefont
  {Weinberg}},\ }\href@noop {} {\bibfield  {journal} {\bibinfo  {journal} {Rev.
  Mod. Phys}\ }\textbf {\bibinfo {volume} {52}},\ \bibinfo {pages} {515}
  (\bibinfo {year} {1980})}\BibitemShut {NoStop}%
\bibitem [{\citenamefont {Bedaque}\ and\ \citenamefont {van
  Kolck}(2002)}]{BK02}%
  \BibitemOpen
  \bibfield  {author} {\bibinfo {author} {\bibfnamefont {P.~F.}\ \bibnamefont
  {Bedaque}}\ and\ \bibinfo {author} {\bibfnamefont {U.}~\bibnamefont {van
  Kolck}},\ }\href {https://doi.org/10.1146/annurev.nucl.52.050102.090637}
  {\bibfield  {journal} {\bibinfo  {journal} {Annu. Rev. Nucl. Part. Sci.}\
  }\textbf {\bibinfo {volume} {52}},\ \bibinfo {pages} {339} (\bibinfo {year}
  {2002})}\BibitemShut {NoStop}%
\bibitem [{\citenamefont {Hammer}\ \emph {et~al.}(2017)\citenamefont {Hammer},
  \citenamefont {Ji},\ and\ \citenamefont {Phillips}}]{HJP17}%
  \BibitemOpen
  \bibfield  {author} {\bibinfo {author} {\bibfnamefont {H.-W.}\ \bibnamefont
  {Hammer}}, \bibinfo {author} {\bibfnamefont {C.}~\bibnamefont {Ji}},\ and\
  \bibinfo {author} {\bibfnamefont {D.~R.}\ \bibnamefont {Phillips}},\ }\href
  {https://doi.org/10.1088/1361-6471/aa83db} {\bibfield  {journal} {\bibinfo
  {journal} {J. Phys. G: Nucl. Part. Phys.}\ }\textbf {\bibinfo {volume}
  {44}},\ \bibinfo {pages} {103002} (\bibinfo {year} {2017})}\BibitemShut
  {NoStop}%
\bibitem [{\citenamefont {Hammer}\ \emph {et~al.}(2020)\citenamefont {Hammer},
  \citenamefont {K\"onig},\ and\ \citenamefont {van Kolck}}]{HKK20}%
  \BibitemOpen
  \bibfield  {author} {\bibinfo {author} {\bibfnamefont {H.-W.}\ \bibnamefont
  {Hammer}}, \bibinfo {author} {\bibfnamefont {S.}~\bibnamefont {K\"onig}},\
  and\ \bibinfo {author} {\bibfnamefont {U.}~\bibnamefont {van Kolck}},\ }\href
  {https://doi.org/10.1103/RevModPhys.92.025004} {\bibfield  {journal}
  {\bibinfo  {journal} {Rev. Mod. Phys.}\ }\textbf {\bibinfo {volume} {92}},\
  \bibinfo {pages} {025004} (\bibinfo {year} {2020})}\BibitemShut {NoStop}%
\bibitem [{\citenamefont {Maris}\ \emph {et~al.}(2021)\citenamefont {Maris},
  \citenamefont {Epelbaum}, \citenamefont {Furnstahl}, \citenamefont {Golak},
  \citenamefont {Hebeler}, \citenamefont {H\"uther}, \citenamefont {Kamada},
  \citenamefont {Krebs}, \citenamefont {Mei\ss{}ner}, \citenamefont {Melendez},
  \citenamefont {Nogga}, \citenamefont {Reinert}, \citenamefont {Roth},
  \citenamefont {Skibi\ifmmode~\acute{n}\else \'{n}\fi{}ski}, \citenamefont
  {Soloviov}, \citenamefont {Topolnicki}, \citenamefont {Vary}, \citenamefont
  {Volkotrub}, \citenamefont {Wita\l{}a},\ and\ \citenamefont
  {Wolfgruber}}]{MEF21}%
  \BibitemOpen
  \bibfield  {author} {\bibinfo {author} {\bibfnamefont {P.}~\bibnamefont
  {Maris}}, \bibinfo {author} {\bibfnamefont {E.}~\bibnamefont {Epelbaum}},
  \bibinfo {author} {\bibfnamefont {R.~J.}\ \bibnamefont {Furnstahl}}, \bibinfo
  {author} {\bibfnamefont {J.}~\bibnamefont {Golak}}, \bibinfo {author}
  {\bibfnamefont {K.}~\bibnamefont {Hebeler}}, \bibinfo {author} {\bibfnamefont
  {T.}~\bibnamefont {H\"uther}}, \bibinfo {author} {\bibfnamefont
  {H.}~\bibnamefont {Kamada}}, \bibinfo {author} {\bibfnamefont
  {H.}~\bibnamefont {Krebs}}, \bibinfo {author} {\bibfnamefont {U.-G.}\
  \bibnamefont {Mei\ss{}ner}}, \bibinfo {author} {\bibfnamefont {J.~A.}\
  \bibnamefont {Melendez}}, \bibinfo {author} {\bibfnamefont {A.}~\bibnamefont
  {Nogga}}, \bibinfo {author} {\bibfnamefont {P.}~\bibnamefont {Reinert}},
  \bibinfo {author} {\bibfnamefont {R.}~\bibnamefont {Roth}}, \bibinfo {author}
  {\bibfnamefont {R.}~\bibnamefont {Skibi\ifmmode~\acute{n}\else
  \'{n}\fi{}ski}}, \bibinfo {author} {\bibfnamefont {V.}~\bibnamefont
  {Soloviov}}, \bibinfo {author} {\bibfnamefont {K.}~\bibnamefont
  {Topolnicki}}, \bibinfo {author} {\bibfnamefont {J.~P.}\ \bibnamefont
  {Vary}}, \bibinfo {author} {\bibfnamefont {Y.}~\bibnamefont {Volkotrub}},
  \bibinfo {author} {\bibfnamefont {H.}~\bibnamefont {Wita\l{}a}},\ and\
  \bibinfo {author} {\bibfnamefont {T.}~\bibnamefont {Wolfgruber}} (\bibinfo
  {collaboration} {LENPIC Collaboration}),\ }\href
  {https://doi.org/10.1103/PhysRevC.103.054001} {\bibfield  {journal} {\bibinfo
   {journal} {Phys. Rev. C}\ }\textbf {\bibinfo {volume} {103}},\ \bibinfo
  {pages} {054001} (\bibinfo {year} {2021})}\BibitemShut {NoStop}%
\bibitem [{\citenamefont {Al-Khalili}(2004)}]{Al04}%
  \BibitemOpen
  \bibfield  {author} {\bibinfo {author} {\bibfnamefont {J.}~\bibnamefont
  {Al-Khalili}},\ }\bibinfo {title} {An introduction to halo nuclei}\ (\bibinfo
   {publisher} {Springer Berlin Heidelberg},\ \bibinfo {year} {2004})\ pp.\
  \bibinfo {pages} {77--112}\BibitemShut {NoStop}%
\bibitem [{\citenamefont {Zhukov}\ \emph {et~al.}(1993)\citenamefont {Zhukov},
  \citenamefont {Danilin}, \citenamefont {Fedorov}, \citenamefont {Bang},
  \citenamefont {Thompson},\ and\ \citenamefont {Vaagen}}]{ZDF93}%
  \BibitemOpen
  \bibfield  {author} {\bibinfo {author} {\bibfnamefont {M.~V.}\ \bibnamefont
  {Zhukov}}, \bibinfo {author} {\bibfnamefont {B.~V.}\ \bibnamefont {Danilin}},
  \bibinfo {author} {\bibfnamefont {D.~V.}\ \bibnamefont {Fedorov}}, \bibinfo
  {author} {\bibfnamefont {J.~M.}\ \bibnamefont {Bang}}, \bibinfo {author}
  {\bibfnamefont {I.~J.}\ \bibnamefont {Thompson}},\ and\ \bibinfo {author}
  {\bibfnamefont {J.~S.}\ \bibnamefont {Vaagen}},\ }\href@noop {} {\bibfield
  {journal} {\bibinfo  {journal} {Phys. Rep.}\ }\textbf {\bibinfo {volume}
  {231}},\ \bibinfo {pages} {151} (\bibinfo {year} {1993})}\BibitemShut
  {NoStop}%
\bibitem [{\citenamefont {Danilin}\ \emph {et~al.}(1998)\citenamefont
  {Danilin}, \citenamefont {Thompson}, \citenamefont {Vaagen},\ and\
  \citenamefont {Zhukov}}]{DTV98}%
  \BibitemOpen
  \bibfield  {author} {\bibinfo {author} {\bibfnamefont {B.~V.}\ \bibnamefont
  {Danilin}}, \bibinfo {author} {\bibfnamefont {I.~J.}\ \bibnamefont
  {Thompson}}, \bibinfo {author} {\bibfnamefont {J.~S.}\ \bibnamefont
  {Vaagen}},\ and\ \bibinfo {author} {\bibfnamefont {M.~V.}\ \bibnamefont
  {Zhukov}},\ }\href@noop {} {\bibfield  {journal} {\bibinfo  {journal} {Nucl.
  Phys. A}\ }\textbf {\bibinfo {volume} {632}},\ \bibinfo {pages} {383}
  (\bibinfo {year} {1998})}\BibitemShut {NoStop}%
\bibitem [{\citenamefont {Descouvemont}\ \emph {et~al.}(2003)\citenamefont
  {Descouvemont}, \citenamefont {Daniel},\ and\ \citenamefont {Baye}}]{DDB03}%
  \BibitemOpen
  \bibfield  {author} {\bibinfo {author} {\bibfnamefont {P.}~\bibnamefont
  {Descouvemont}}, \bibinfo {author} {\bibfnamefont {C.}~\bibnamefont
  {Daniel}},\ and\ \bibinfo {author} {\bibfnamefont {D.}~\bibnamefont {Baye}},\
  }\href@noop {} {\bibfield  {journal} {\bibinfo  {journal} {Phys. Rev. C}\
  }\textbf {\bibinfo {volume} {67}},\ \bibinfo {pages} {044309} (\bibinfo
  {year} {2003})}\BibitemShut {NoStop}%
\bibitem [{\citenamefont {Baye}\ \emph {et~al.}(2009)\citenamefont {Baye},
  \citenamefont {Capel}, \citenamefont {Descouvemont},\ and\ \citenamefont
  {Suzuki}}]{BCD09}%
  \BibitemOpen
  \bibfield  {author} {\bibinfo {author} {\bibfnamefont {D.}~\bibnamefont
  {Baye}}, \bibinfo {author} {\bibfnamefont {P.}~\bibnamefont {Capel}},
  \bibinfo {author} {\bibfnamefont {P.}~\bibnamefont {Descouvemont}},\ and\
  \bibinfo {author} {\bibfnamefont {Y.}~\bibnamefont {Suzuki}},\ }\href@noop {}
  {\bibfield  {journal} {\bibinfo  {journal} {Phys. Rev. C}\ }\textbf {\bibinfo
  {volume} {79}},\ \bibinfo {pages} {024607} (\bibinfo {year}
  {2009})}\BibitemShut {NoStop}%
\bibitem [{\citenamefont {Rodr\'{\i}guez-Gallardo}\ \emph
  {et~al.}(2008)\citenamefont {Rodr\'{\i}guez-Gallardo}, \citenamefont {Arias},
  \citenamefont {G\'{o}mez-Camacho}, \citenamefont {Johnson}, \citenamefont
  {Moro}, \citenamefont {Thompson},\ and\ \citenamefont {Tostevin}}]{RAG08}%
  \BibitemOpen
  \bibfield  {author} {\bibinfo {author} {\bibfnamefont {M.}~\bibnamefont
  {Rodr\'{\i}guez-Gallardo}}, \bibinfo {author} {\bibfnamefont {J.~M.}\
  \bibnamefont {Arias}}, \bibinfo {author} {\bibfnamefont {J.}~\bibnamefont
  {G\'{o}mez-Camacho}}, \bibinfo {author} {\bibfnamefont {R.~C.}\ \bibnamefont
  {Johnson}}, \bibinfo {author} {\bibfnamefont {A.~M.}\ \bibnamefont {Moro}},
  \bibinfo {author} {\bibfnamefont {I.~J.}\ \bibnamefont {Thompson}},\ and\
  \bibinfo {author} {\bibfnamefont {J.~A.}\ \bibnamefont {Tostevin}},\
  }\href@noop {} {\bibfield  {journal} {\bibinfo  {journal} {Phys. Rev. C}\
  }\textbf {\bibinfo {volume} {77}},\ \bibinfo {pages} {064609} (\bibinfo
  {year} {2008})}\BibitemShut {NoStop}%
\bibitem [{\citenamefont {Ji}\ \emph {et~al.}(2014)\citenamefont {Ji},
  \citenamefont {Elster},\ and\ \citenamefont {Phillips}}]{JEP14}%
  \BibitemOpen
  \bibfield  {author} {\bibinfo {author} {\bibfnamefont {C.}~\bibnamefont
  {Ji}}, \bibinfo {author} {\bibfnamefont {C.}~\bibnamefont {Elster}},\ and\
  \bibinfo {author} {\bibfnamefont {D.~R.}\ \bibnamefont {Phillips}},\ }\href
  {https://doi.org/10.1103/PhysRevC.90.044004} {\bibfield  {journal} {\bibinfo
  {journal} {Phys. Rev. C}\ }\textbf {\bibinfo {volume} {90}},\ \bibinfo
  {pages} {044004} (\bibinfo {year} {2014})}\BibitemShut {NoStop}%
\bibitem [{\citenamefont {Faddeev}(1961)}]{Fa61}%
  \BibitemOpen
  \bibfield  {author} {\bibinfo {author} {\bibfnamefont {L.}~\bibnamefont
  {Faddeev}},\ }\href@noop {} {\bibfield  {journal} {\bibinfo  {journal} {Zh.
  Eksp. Teor. Fiz}\ }\textbf {\bibinfo {volume} {39}},\ \bibinfo {pages} {1014}
  (\bibinfo {year} {1961})}\BibitemShut {NoStop}%
\bibitem [{\citenamefont {Filandri}\ \emph {et~al.}(2020)\citenamefont
  {Filandri}, \citenamefont {Andreatta}, \citenamefont {Manzata}, \citenamefont
  {Ji}, \citenamefont {Leidemann},\ and\ \citenamefont {Orlandini}}]{FA20}%
  \BibitemOpen
  \bibfield  {author} {\bibinfo {author} {\bibfnamefont {E.}~\bibnamefont
  {Filandri}}, \bibinfo {author} {\bibfnamefont {P.}~\bibnamefont {Andreatta}},
  \bibinfo {author} {\bibfnamefont {C.~A.}\ \bibnamefont {Manzata}}, \bibinfo
  {author} {\bibfnamefont {C.}~\bibnamefont {Ji}}, \bibinfo {author}
  {\bibfnamefont {W.}~\bibnamefont {Leidemann}},\ and\ \bibinfo {author}
  {\bibfnamefont {G.}~\bibnamefont {Orlandini}},\ }\href
  {https://doi.org/10.21468/SciPostPhysProc.3.034} {\bibfield  {journal}
  {\bibinfo  {journal} {SciPost Phys. Proc.}\ ,\ \bibinfo {pages} {034}}
  (\bibinfo {year} {2020})}\BibitemShut {NoStop}%
\bibitem [{\citenamefont {Filandri}(2022)}]{F22}%
  \BibitemOpen
  \bibfield  {author} {\bibinfo {author} {\bibfnamefont {E.}~\bibnamefont
  {Filandri}},\ }\emph {\bibinfo {title} {Effective field theory description of
  $\alpha$-cluster nuclei:The $^{9}$Be ground state and $^{9}$Be
  photodisintegration}},\ \href@noop {} {Ph.D. thesis},\ \bibinfo  {school}
  {University of Trento} (\bibinfo {year} {2022})\BibitemShut {NoStop}%
\bibitem [{\citenamefont {Capitani}\ \emph {et~al.}()\citenamefont {Capitani},
  \citenamefont {Filandri}, \citenamefont {Ji},\ and\ \citenamefont
  {Leidemann}}]{CFJ24}%
  \BibitemOpen
  \bibfield  {author} {\bibinfo {author} {\bibfnamefont {Y.}~\bibnamefont
  {Capitani}}, \bibinfo {author} {\bibfnamefont {E.}~\bibnamefont {Filandri}},
  \bibinfo {author} {\bibfnamefont {C.}~\bibnamefont {Ji}},\ and\ \bibinfo
  {author} {\bibfnamefont {W.}~\bibnamefont {Leidemann}},\ }\href@noop {}
  {\bibinfo  {journal} {in preparation}\ }\BibitemShut {NoStop}%
\bibitem [{\citenamefont {Theeten}\ \emph {et~al.}(2005)\citenamefont
  {Theeten}, \citenamefont {Baye},\ and\ \citenamefont {Descouvemont}}]{TBD05}%
  \BibitemOpen
\bibfield  {journal} {  }\bibfield  {author} {\bibinfo {author} {\bibfnamefont
  {M.}~\bibnamefont {Theeten}}, \bibinfo {author} {\bibfnamefont
  {D.}~\bibnamefont {Baye}},\ and\ \bibinfo {author} {\bibfnamefont
  {P.}~\bibnamefont {Descouvemont}},\ }\href@noop {} {\bibfield  {journal}
  {\bibinfo  {journal} {Nucl. Phys. A}\ }\textbf {\bibinfo {volume} {753}},\
  \bibinfo {pages} {233} (\bibinfo {year} {2005})}\BibitemShut {NoStop}%
\bibitem [{\citenamefont {Kukulin}\ and\ \citenamefont
  {Pomerantsev}(1978)}]{KP78}%
  \BibitemOpen
  \bibfield  {author} {\bibinfo {author} {\bibfnamefont {V.~I.}\ \bibnamefont
  {Kukulin}}\ and\ \bibinfo {author} {\bibfnamefont {V.~N.}\ \bibnamefont
  {Pomerantsev}},\ }\href@noop {} {\bibfield  {journal} {\bibinfo  {journal}
  {Ann. Phys.}\ }\textbf {\bibinfo {volume} {111}},\ \bibinfo {pages} {330}
  (\bibinfo {year} {1978})}\BibitemShut {NoStop}%
\bibitem [{\citenamefont {Raynal}\ and\ \citenamefont {Revai}(1970)}]{RR70}%
  \BibitemOpen
  \bibfield  {author} {\bibinfo {author} {\bibfnamefont {J.}~\bibnamefont
  {Raynal}}\ and\ \bibinfo {author} {\bibfnamefont {J.}~\bibnamefont {Revai}},\
  }\href@noop {} {\bibfield  {journal} {\bibinfo  {journal} {Nuovo Cim. A}\
  }\textbf {\bibinfo {volume} {39}},\ \bibinfo {pages} {612} (\bibinfo {year}
  {1970})}\BibitemShut {NoStop}%
\bibitem [{\citenamefont {Baye}(2015)}]{Ba15}%
  \BibitemOpen
  \bibfield  {author} {\bibinfo {author} {\bibfnamefont {D.}~\bibnamefont
  {Baye}},\ }\href
  {https://doi.org/https://doi.org/10.1016/j.physrep.2014.11.006} {\bibfield
  {journal} {\bibinfo  {journal} {Phys. Rep.}\ }\textbf {\bibinfo {volume}
  {565}},\ \bibinfo {pages} {1} (\bibinfo {year} {2015})}\BibitemShut {NoStop}%
\bibitem [{\citenamefont {Wigner}(1955)}]{Wi55}%
  \BibitemOpen
  \bibfield  {author} {\bibinfo {author} {\bibfnamefont {E.~P.}\ \bibnamefont
  {Wigner}},\ }\href {https://doi.org/10.1103/PhysRev.98.145} {\bibfield
  {journal} {\bibinfo  {journal} {Phys. Rev.}\ }\textbf {\bibinfo {volume}
  {98}},\ \bibinfo {pages} {145} (\bibinfo {year} {1955})}\BibitemShut
  {NoStop}%
\bibitem [{\citenamefont {Joachain}(1975)}]{Jo75}%
  \BibitemOpen
  \bibfield  {author} {\bibinfo {author} {\bibfnamefont {C.}~\bibnamefont
  {Joachain}},\ }\href {https://books.google.com.co/books?id=Zs3vAAAAMAAJ}
  {\emph {\bibinfo {title} {Quantum Collision Theory}}}\ (\bibinfo  {publisher}
  {North-Holland Publishing Company},\ \bibinfo {year} {1975})\BibitemShut
  {NoStop}%
\bibitem [{\citenamefont {Bethe}(1949)}]{Be49}%
  \BibitemOpen
  \bibfield  {author} {\bibinfo {author} {\bibfnamefont {H.~A.}\ \bibnamefont
  {Bethe}},\ }\href {https://doi.org/10.1103/PhysRev.76.38} {\bibfield
  {journal} {\bibinfo  {journal} {Phys. Rev.}\ }\textbf {\bibinfo {volume}
  {76}},\ \bibinfo {pages} {38} (\bibinfo {year} {1949})}\BibitemShut {NoStop}%
\bibitem [{\citenamefont {Blatt}\ and\ \citenamefont {Jackson}(1949)}]{BJ49}%
  \BibitemOpen
  \bibfield  {author} {\bibinfo {author} {\bibfnamefont {J.~M.}\ \bibnamefont
  {Blatt}}\ and\ \bibinfo {author} {\bibfnamefont {J.~D.}\ \bibnamefont
  {Jackson}},\ }\href {https://doi.org/10.1103/PhysRev.76.18} {\bibfield
  {journal} {\bibinfo  {journal} {Phys. Rev.}\ }\textbf {\bibinfo {volume}
  {76}},\ \bibinfo {pages} {18} (\bibinfo {year} {1949})}\BibitemShut {NoStop}%
\bibitem [{\citenamefont {Buck}\ \emph {et~al.}(1977)\citenamefont {Buck},
  \citenamefont {Friedrich},\ and\ \citenamefont {Wheatley}}]{BFW77}%
  \BibitemOpen
  \bibfield  {author} {\bibinfo {author} {\bibfnamefont {B.}~\bibnamefont
  {Buck}}, \bibinfo {author} {\bibfnamefont {H.}~\bibnamefont {Friedrich}},\
  and\ \bibinfo {author} {\bibfnamefont {C.}~\bibnamefont {Wheatley}},\
  }\href@noop {} {\bibfield  {journal} {\bibinfo  {journal} {Nucl. Phys. A}\
  }\textbf {\bibinfo {volume} {275}},\ \bibinfo {pages} {246} (\bibinfo {year}
  {1977})}\BibitemShut {NoStop}%
\bibitem [{\citenamefont {Baye}(1987)}]{Ba87}%
  \BibitemOpen
  \bibfield  {author} {\bibinfo {author} {\bibfnamefont {D.}~\bibnamefont
  {Baye}},\ }\href {https://doi.org/10.1103/PhysRevLett.58.2738} {\bibfield
  {journal} {\bibinfo  {journal} {Phys. Rev. Lett.}\ }\textbf {\bibinfo
  {volume} {58}},\ \bibinfo {pages} {2738} (\bibinfo {year}
  {1987})}\BibitemShut {NoStop}%
\bibitem [{\citenamefont {Capitani}(2024)}]{Yl24}%
  \BibitemOpen
  \bibfield  {author} {\bibinfo {author} {\bibfnamefont {Y.}~\bibnamefont
  {Capitani}},\ }\emph {\bibinfo {title} {Cluster Effective Field Theory
  calculation of electromagnetic breakup reactions with the Lorentz Integral
  Transform method}},\ \href@noop {} {Ph.D. thesis},\ \bibinfo  {school}
  {University of Trento} (\bibinfo {year} {2024})\BibitemShut {NoStop}%
\bibitem [{\citenamefont {Suzuki}\ \emph {et~al.}(2003)\citenamefont {Suzuki},
  \citenamefont {Lovas}, \citenamefont {Yabana},\ and\ \citenamefont
  {Varga}}]{SLY03}%
  \BibitemOpen
  \bibfield  {author} {\bibinfo {author} {\bibfnamefont {Y.}~\bibnamefont
  {Suzuki}}, \bibinfo {author} {\bibfnamefont {R.~G.}\ \bibnamefont {Lovas}},
  \bibinfo {author} {\bibfnamefont {K.}~\bibnamefont {Yabana}},\ and\ \bibinfo
  {author} {\bibfnamefont {K.}~\bibnamefont {Varga}},\ }\href@noop {} {\emph
  {\bibinfo {title} {Structure and Reactions of Light Exotic Nuclei}}}\
  (\bibinfo  {publisher} {Taylor \& Francis, London},\ \bibinfo {year}
  {2003})\BibitemShut {NoStop}%
\bibitem [{\citenamefont {Golak}\ \emph {et~al.}(2005)\citenamefont {Golak},
  \citenamefont {Skibiński}, \citenamefont {Witała}, \citenamefont
  {Glöckle}, \citenamefont {Nogga},\ and\ \citenamefont {Kamada}}]{GSW05}%
  \BibitemOpen
  \bibfield  {author} {\bibinfo {author} {\bibfnamefont {J.}~\bibnamefont
  {Golak}}, \bibinfo {author} {\bibfnamefont {R.}~\bibnamefont {Skibiński}},
  \bibinfo {author} {\bibfnamefont {H.}~\bibnamefont {Witała}}, \bibinfo
  {author} {\bibfnamefont {W.}~\bibnamefont {Glöckle}}, \bibinfo {author}
  {\bibfnamefont {A.}~\bibnamefont {Nogga}},\ and\ \bibinfo {author}
  {\bibfnamefont {H.}~\bibnamefont {Kamada}},\ }\href
  {https://doi.org/https://doi.org/10.1016/j.physrep.2005.04.005} {\bibfield
  {journal} {\bibinfo  {journal} {Phys. Rep.}\ }\textbf {\bibinfo {volume}
  {415}},\ \bibinfo {pages} {89} (\bibinfo {year} {2005})}\BibitemShut
  {NoStop}%
\bibitem [{\citenamefont {Descouvemont}\ \emph {et~al.}(2006)\citenamefont
  {Descouvemont}, \citenamefont {Tursunov},\ and\ \citenamefont
  {Baye}}]{DTB06}%
  \BibitemOpen
  \bibfield  {author} {\bibinfo {author} {\bibfnamefont {P.}~\bibnamefont
  {Descouvemont}}, \bibinfo {author} {\bibfnamefont {E.~M.}\ \bibnamefont
  {Tursunov}},\ and\ \bibinfo {author} {\bibfnamefont {D.}~\bibnamefont
  {Baye}},\ }\href@noop {} {\bibfield  {journal} {\bibinfo  {journal} {Nucl.
  Phys. A}\ }\textbf {\bibinfo {volume} {765}},\ \bibinfo {pages} {370}
  (\bibinfo {year} {2006})}\BibitemShut {NoStop}%
\bibitem [{\citenamefont {Aoyama}\ \emph {et~al.}(2006)\citenamefont {Aoyama},
  \citenamefont {Myo}, \citenamefont {Kat\={o}},\ and\ \citenamefont
  {Ikeda}}]{AMK06}%
  \BibitemOpen
  \bibfield  {author} {\bibinfo {author} {\bibfnamefont {S.}~\bibnamefont
  {Aoyama}}, \bibinfo {author} {\bibfnamefont {T.}~\bibnamefont {Myo}},
  \bibinfo {author} {\bibfnamefont {K.}~\bibnamefont {Kat\={o}}},\ and\
  \bibinfo {author} {\bibfnamefont {K.}~\bibnamefont {Ikeda}},\ }\href@noop {}
  {\bibfield  {journal} {\bibinfo  {journal} {Prog. Theor. Phys.}\ }\textbf
  {\bibinfo {volume} {116}},\ \bibinfo {pages} {1} (\bibinfo {year}
  {2006})}\BibitemShut {NoStop}%
\bibitem [{\citenamefont {Descouvemont}\ \emph {et~al.}(2012)\citenamefont
  {Descouvemont}, \citenamefont {Pinilla},\ and\ \citenamefont {Baye}}]{DPB12}%
  \BibitemOpen
  \bibfield  {author} {\bibinfo {author} {\bibfnamefont {P.}~\bibnamefont
  {Descouvemont}}, \bibinfo {author} {\bibfnamefont {E.~C.}\ \bibnamefont
  {Pinilla}},\ and\ \bibinfo {author} {\bibfnamefont {D.}~\bibnamefont
  {Baye}},\ }\href {https://doi.org/10.1143/PTPS.196.1} {\bibfield  {journal}
  {\bibinfo  {journal} {Prog. Theor. Phys. Suppl.}\ }\textbf {\bibinfo {volume}
  {196}},\ \bibinfo {pages} {1} (\bibinfo {year} {2012})}\BibitemShut {NoStop}%
\bibitem [{\citenamefont {Carlson}\ and\ \citenamefont
  {Schiavilla}(1998)}]{CS98}%
  \BibitemOpen
  \bibfield  {author} {\bibinfo {author} {\bibfnamefont {J.}~\bibnamefont
  {Carlson}}\ and\ \bibinfo {author} {\bibfnamefont {R.}~\bibnamefont
  {Schiavilla}},\ }\href {https://doi.org/10.1103/RevModPhys.70.743} {\bibfield
   {journal} {\bibinfo  {journal} {Rev. Mod. Phys.}\ }\textbf {\bibinfo
  {volume} {70}},\ \bibinfo {pages} {743} (\bibinfo {year} {1998})}\BibitemShut
  {NoStop}%
\bibitem [{\citenamefont {Roggero}(2020)}]{Ro20}%
  \BibitemOpen
  \bibfield  {author} {\bibinfo {author} {\bibfnamefont {A.}~\bibnamefont
  {Roggero}},\ }\href {https://doi.org/10.1103/PhysRevA.102.022409} {\bibfield
  {journal} {\bibinfo  {journal} {Phys. Rev. A}\ }\textbf {\bibinfo {volume}
  {102}},\ \bibinfo {pages} {022409} (\bibinfo {year} {2020})}\BibitemShut
  {NoStop}%
\bibitem [{\citenamefont {Rodr\'{\i}guez-Gallardo}\ \emph
  {et~al.}(2005)\citenamefont {Rodr\'{\i}guez-Gallardo}, \citenamefont {Arias},
  \citenamefont {G\'omez-Camacho}, \citenamefont {Moro}, \citenamefont
  {Thompson},\ and\ \citenamefont {Tostevin}}]{RAG05}%
  \BibitemOpen
  \bibfield  {author} {\bibinfo {author} {\bibfnamefont {M.}~\bibnamefont
  {Rodr\'{\i}guez-Gallardo}}, \bibinfo {author} {\bibfnamefont {J.~M.}\
  \bibnamefont {Arias}}, \bibinfo {author} {\bibfnamefont {J.}~\bibnamefont
  {G\'omez-Camacho}}, \bibinfo {author} {\bibfnamefont {A.~M.}\ \bibnamefont
  {Moro}}, \bibinfo {author} {\bibfnamefont {I.~J.}\ \bibnamefont {Thompson}},\
  and\ \bibinfo {author} {\bibfnamefont {J.~A.}\ \bibnamefont {Tostevin}},\
  }\href {https://doi.org/10.1103/PhysRevC.72.024007} {\bibfield  {journal}
  {\bibinfo  {journal} {Phys. Rev. C}\ }\textbf {\bibinfo {volume} {72}},\
  \bibinfo {pages} {024007} (\bibinfo {year} {2005})}\BibitemShut {NoStop}%
\bibitem [{\citenamefont {Pinilla}\ \emph {et~al.}(2011)\citenamefont
  {Pinilla}, \citenamefont {Baye}, \citenamefont {Descouvemont}, \citenamefont
  {Horiuchi},\ and\ \citenamefont {Suzuki}}]{PBD11}%
  \BibitemOpen
  \bibfield  {author} {\bibinfo {author} {\bibfnamefont {E.~C.}\ \bibnamefont
  {Pinilla}}, \bibinfo {author} {\bibfnamefont {D.}~\bibnamefont {Baye}},
  \bibinfo {author} {\bibfnamefont {P.}~\bibnamefont {Descouvemont}}, \bibinfo
  {author} {\bibfnamefont {W.}~\bibnamefont {Horiuchi}},\ and\ \bibinfo
  {author} {\bibfnamefont {Y.}~\bibnamefont {Suzuki}},\ }\href
  {https://doi.org/DOI: 10.1016/j.nuclphysa.2011.06.030} {\bibfield  {journal}
  {\bibinfo  {journal} {Nucl. Phys. A}\ }\textbf {\bibinfo {volume} {865}},\
  \bibinfo {pages} {43 } (\bibinfo {year} {2011})}\BibitemShut {NoStop}%
\bibitem [{\citenamefont {Casal}\ \emph {et~al.}(2020)\citenamefont {Casal},
  \citenamefont {Singh}, \citenamefont {Fortunato}, \citenamefont {Horiuchi},\
  and\ \citenamefont {Vitturi}}]{CSF20}%
  \BibitemOpen
  \bibfield  {author} {\bibinfo {author} {\bibfnamefont {J.}~\bibnamefont
  {Casal}}, \bibinfo {author} {\bibfnamefont {J.}~\bibnamefont {Singh}},
  \bibinfo {author} {\bibfnamefont {L.}~\bibnamefont {Fortunato}}, \bibinfo
  {author} {\bibfnamefont {W.}~\bibnamefont {Horiuchi}},\ and\ \bibinfo
  {author} {\bibfnamefont {A.}~\bibnamefont {Vitturi}},\ }\href
  {https://doi.org/10.1103/PhysRevC.102.064627} {\bibfield  {journal} {\bibinfo
   {journal} {Phys. Rev. C}\ }\textbf {\bibinfo {volume} {102}},\ \bibinfo
  {pages} {064627} (\bibinfo {year} {2020})}\BibitemShut {NoStop}%
\bibitem [{\citenamefont {Leo}(2012)}]{Le12}%
  \BibitemOpen
  \bibfield  {author} {\bibinfo {author} {\bibfnamefont {W.~R.}\ \bibnamefont
  {Leo}},\ }\href@noop {} {\emph {\bibinfo {title} {Techniques for Nuclear and
  Particle Physics Experiments: A How-to Approach}}}\ (\bibinfo  {publisher}
  {Springer Berlin Heidelberg},\ \bibinfo {year} {2012})\BibitemShut {NoStop}%
\bibitem [{\citenamefont {Haxton}\ \emph {et~al.}(2005)\citenamefont {Haxton},
  \citenamefont {Nollett},\ and\ \citenamefont {Zurek}}]{HN05}%
  \BibitemOpen
  \bibfield  {author} {\bibinfo {author} {\bibfnamefont {W.~C.}\ \bibnamefont
  {Haxton}}, \bibinfo {author} {\bibfnamefont {K.~M.}\ \bibnamefont
  {Nollett}},\ and\ \bibinfo {author} {\bibfnamefont {K.~M.}\ \bibnamefont
  {Zurek}},\ }\href {https://doi.org/10.1103/PhysRevC.72.065501} {\bibfield
  {journal} {\bibinfo  {journal} {Phys. Rev. C}\ }\textbf {\bibinfo {volume}
  {72}},\ \bibinfo {pages} {065501} (\bibinfo {year} {2005})}\BibitemShut
  {NoStop}%
\bibitem [{\citenamefont {Orlandini}\ and\ \citenamefont {Turro}(2017)}]{OT17}%
  \BibitemOpen
  \bibfield  {author} {\bibinfo {author} {\bibfnamefont {G.}~\bibnamefont
  {Orlandini}}\ and\ \bibinfo {author} {\bibfnamefont {F.}~\bibnamefont
  {Turro}},\ }\href@noop {} {\bibfield  {journal} {\bibinfo  {journal}
  {Few-Body Syst.}\ }\textbf {\bibinfo {volume} {58}} (\bibinfo {year}
  {2017})}\BibitemShut {NoStop}%
\bibitem [{\citenamefont {Efros}\ \emph {et~al.}(1998)\citenamefont {Efros},
  \citenamefont {Leidemann},\ and\ \citenamefont {Orlandini}}]{ELO98}%
  \BibitemOpen
  \bibfield  {author} {\bibinfo {author} {\bibfnamefont {V.}~\bibnamefont
  {Efros}}, \bibinfo {author} {\bibfnamefont {W.}~\bibnamefont {Leidemann}},\
  and\ \bibinfo {author} {\bibfnamefont {G.}~\bibnamefont {Orlandini}},\ }\href
  {https://doi.org/https://doi.org/10.1016/S0375-9474(98)00086-4} {\bibfield
  {journal} {\bibinfo  {journal} {Nucl. Phys. A}\ }\textbf {\bibinfo {volume}
  {631}},\ \bibinfo {pages} {658} (\bibinfo {year} {1998})}\BibitemShut
  {NoStop}%
\bibitem [{\citenamefont {Efros}\ \emph {et~al.}(2019)\citenamefont {Efros},
  \citenamefont {Leidemann},\ and\ \citenamefont {Shalamova}}]{ELS19}%
  \BibitemOpen
  \bibfield  {author} {\bibinfo {author} {\bibfnamefont {V.~D.}\ \bibnamefont
  {Efros}}, \bibinfo {author} {\bibfnamefont {W.}~\bibnamefont {Leidemann}},\
  and\ \bibinfo {author} {\bibfnamefont {V.~Y.}\ \bibnamefont {Shalamova}},\
  }\href@noop {} {\bibfield  {journal} {\bibinfo  {journal} {Few-Body Syst.}\
  }\textbf {\bibinfo {volume} {60}} (\bibinfo {year} {2019})}\BibitemShut
  {NoStop}%
\bibitem [{\citenamefont {Chen}\ \emph {et~al.}(2008)\citenamefont {Chen},
  \citenamefont {Howell}, \citenamefont {Carman}, \citenamefont {Gibbs},
  \citenamefont {Gibson}, \citenamefont {Hussein}, \citenamefont {Kiser},
  \citenamefont {Mertens}, \citenamefont {Moore}, \citenamefont {Morris},
  \citenamefont {Obst}, \citenamefont {Pasyuk}, \citenamefont {Roper},
  \citenamefont {Salinas}, \citenamefont {Setze}, \citenamefont {Slaus},
  \citenamefont {Sterbenz}, \citenamefont {Tornow}, \citenamefont {Walter},
  \citenamefont {Whiteley},\ and\ \citenamefont {Whitton}}]{CC08}%
  \BibitemOpen
  \bibfield  {author} {\bibinfo {author} {\bibfnamefont {Q.}~\bibnamefont
  {Chen}}, \bibinfo {author} {\bibfnamefont {C.~R.}\ \bibnamefont {Howell}},
  \bibinfo {author} {\bibfnamefont {T.~S.}\ \bibnamefont {Carman}}, \bibinfo
  {author} {\bibfnamefont {W.~R.}\ \bibnamefont {Gibbs}}, \bibinfo {author}
  {\bibfnamefont {B.~F.}\ \bibnamefont {Gibson}}, \bibinfo {author}
  {\bibfnamefont {A.}~\bibnamefont {Hussein}}, \bibinfo {author} {\bibfnamefont
  {M.~R.}\ \bibnamefont {Kiser}}, \bibinfo {author} {\bibfnamefont
  {G.}~\bibnamefont {Mertens}}, \bibinfo {author} {\bibfnamefont {C.~F.}\
  \bibnamefont {Moore}}, \bibinfo {author} {\bibfnamefont {C.}~\bibnamefont
  {Morris}}, \bibinfo {author} {\bibfnamefont {A.}~\bibnamefont {Obst}},
  \bibinfo {author} {\bibfnamefont {E.}~\bibnamefont {Pasyuk}}, \bibinfo
  {author} {\bibfnamefont {C.~D.}\ \bibnamefont {Roper}}, \bibinfo {author}
  {\bibfnamefont {F.}~\bibnamefont {Salinas}}, \bibinfo {author} {\bibfnamefont
  {H.~R.}\ \bibnamefont {Setze}}, \bibinfo {author} {\bibfnamefont
  {I.}~\bibnamefont {Slaus}}, \bibinfo {author} {\bibfnamefont
  {S.}~\bibnamefont {Sterbenz}}, \bibinfo {author} {\bibfnamefont
  {W.}~\bibnamefont {Tornow}}, \bibinfo {author} {\bibfnamefont {R.~L.}\
  \bibnamefont {Walter}}, \bibinfo {author} {\bibfnamefont {C.~R.}\
  \bibnamefont {Whiteley}},\ and\ \bibinfo {author} {\bibfnamefont
  {M.}~\bibnamefont {Whitton}},\ }\href
  {https://doi.org/10.1103/PhysRevC.77.054002} {\bibfield  {journal} {\bibinfo
  {journal} {Phys. Rev. C}\ }\textbf {\bibinfo {volume} {77}},\ \bibinfo
  {pages} {054002} (\bibinfo {year} {2008})}\BibitemShut {NoStop}%
\bibitem [{\citenamefont {Malone}\ \emph {et~al.}(2022)\citenamefont {Malone},
  \citenamefont {Crowell}, \citenamefont {Cumberbatch}, \citenamefont {Fallin},
  \citenamefont {Friesen}, \citenamefont {Howell}, \citenamefont {Malone},
  \citenamefont {Ticehurst}, \citenamefont {Tornow}, \citenamefont {Markoff},
  \citenamefont {Crowe},\ and\ \citenamefont {Wita?a}}]{Ma22}%
  \BibitemOpen
  \bibfield  {author} {\bibinfo {author} {\bibfnamefont {R.}~\bibnamefont
  {Malone}}, \bibinfo {author} {\bibfnamefont {A.}~\bibnamefont {Crowell}},
  \bibinfo {author} {\bibfnamefont {L.}~\bibnamefont {Cumberbatch}}, \bibinfo
  {author} {\bibfnamefont {B.}~\bibnamefont {Fallin}}, \bibinfo {author}
  {\bibfnamefont {F.}~\bibnamefont {Friesen}}, \bibinfo {author} {\bibfnamefont
  {C.}~\bibnamefont {Howell}}, \bibinfo {author} {\bibfnamefont
  {C.}~\bibnamefont {Malone}}, \bibinfo {author} {\bibfnamefont
  {D.}~\bibnamefont {Ticehurst}}, \bibinfo {author} {\bibfnamefont
  {W.}~\bibnamefont {Tornow}}, \bibinfo {author} {\bibfnamefont
  {D.}~\bibnamefont {Markoff}}, \bibinfo {author} {\bibfnamefont
  {B.}~\bibnamefont {Crowe}},\ and\ \bibinfo {author} {\bibfnamefont
  {H.}~\bibnamefont {Wita?a}},\ }\href
  {https://doi.org/https://doi.org/10.1016/j.physletb.2022.137557} {\bibfield
  {journal} {\bibinfo  {journal} {Phys. Lett. B}\ }\textbf {\bibinfo {volume}
  {835}},\ \bibinfo {pages} {137557} (\bibinfo {year} {2022})}\BibitemShut
  {NoStop}%
\bibitem [{\citenamefont {Arndt}\ \emph {et~al.}(1973)\citenamefont {Arndt},
  \citenamefont {Long},\ and\ \citenamefont {Roper}}]{ALR73}%
  \BibitemOpen
  \bibfield  {author} {\bibinfo {author} {\bibfnamefont {R.~A.}\ \bibnamefont
  {Arndt}}, \bibinfo {author} {\bibfnamefont {D.~D.}\ \bibnamefont {Long}},\
  and\ \bibinfo {author} {\bibfnamefont {L.}~\bibnamefont {Roper}},\ }\href
  {https://doi.org/https://doi.org/10.1016/0375-9474(73)90837-3} {\bibfield
  {journal} {\bibinfo  {journal} {Nucl. Phys. A}\ }\textbf {\bibinfo {volume}
  {209}},\ \bibinfo {pages} {429} (\bibinfo {year} {1973})}\BibitemShut
  {NoStop}%
\bibitem [{\citenamefont {Wiringa}\ \emph {et~al.}(1995)\citenamefont
  {Wiringa}, \citenamefont {Stoks},\ and\ \citenamefont {Schiavilla}}]{WSS95}%
  \BibitemOpen
  \bibfield  {author} {\bibinfo {author} {\bibfnamefont {R.~B.}\ \bibnamefont
  {Wiringa}}, \bibinfo {author} {\bibfnamefont {V.~G.~J.}\ \bibnamefont
  {Stoks}},\ and\ \bibinfo {author} {\bibfnamefont {R.}~\bibnamefont
  {Schiavilla}},\ }\href {https://doi.org/10.1103/PhysRevC.51.38} {\bibfield
  {journal} {\bibinfo  {journal} {Phys. Rev. C}\ }\textbf {\bibinfo {volume}
  {51}},\ \bibinfo {pages} {38} (\bibinfo {year} {1995})}\BibitemShut {NoStop}%
\bibitem [{\citenamefont {Ram\'irez~Su\'arez}\ and\ \citenamefont
  {Sparenberg}(2013)}]{RS13}%
  \BibitemOpen
  \bibfield  {author} {\bibinfo {author} {\bibfnamefont {O.~L.}\ \bibnamefont
  {Ram\'irez~Su\'arez}}\ and\ \bibinfo {author} {\bibfnamefont {J.-M.}\
  \bibnamefont {Sparenberg}},\ }\href
  {https://doi.org/10.1103/PhysRevC.88.014601} {\bibfield  {journal} {\bibinfo
  {journal} {Phys. Rev. C}\ }\textbf {\bibinfo {volume} {88}},\ \bibinfo
  {pages} {014601} (\bibinfo {year} {2013})}\BibitemShut {NoStop}%
\bibitem [{\citenamefont {Lacroix}\ \emph {et~al.}(2012)\citenamefont
  {Lacroix}, \citenamefont {Semay},\ and\ \citenamefont {Buisseret}}]{LSB12}%
  \BibitemOpen
  \bibfield  {author} {\bibinfo {author} {\bibfnamefont {G.}~\bibnamefont
  {Lacroix}}, \bibinfo {author} {\bibfnamefont {C.}~\bibnamefont {Semay}},\
  and\ \bibinfo {author} {\bibfnamefont {F.}~\bibnamefont {Buisseret}},\ }\href
  {https://doi.org/10.1103/PhysRevE.86.026705} {\bibfield  {journal} {\bibinfo
  {journal} {Phys. Rev. E}\ }\textbf {\bibinfo {volume} {86}},\ \bibinfo
  {pages} {026705} (\bibinfo {year} {2012})}\BibitemShut {NoStop}%
\bibitem [{\citenamefont {Kanada}\ \emph {et~al.}(1979)\citenamefont {Kanada},
  \citenamefont {Kaneko}, \citenamefont {Nagata},\ and\ \citenamefont
  {Nomoto}}]{KKN79}%
  \BibitemOpen
  \bibfield  {author} {\bibinfo {author} {\bibfnamefont {H.}~\bibnamefont
  {Kanada}}, \bibinfo {author} {\bibfnamefont {T.}~\bibnamefont {Kaneko}},
  \bibinfo {author} {\bibfnamefont {S.}~\bibnamefont {Nagata}},\ and\ \bibinfo
  {author} {\bibfnamefont {M.}~\bibnamefont {Nomoto}},\ }\href@noop {}
  {\bibfield  {journal} {\bibinfo  {journal} {Prog. Theor. Phys.}\ }\textbf
  {\bibinfo {volume} {61}},\ \bibinfo {pages} {1327} (\bibinfo {year}
  {1979})}\BibitemShut {NoStop}%
\bibitem [{\citenamefont {Sick}(2008)}]{Si08}%
  \BibitemOpen
  \bibfield  {author} {\bibinfo {author} {\bibfnamefont {I.}~\bibnamefont
  {Sick}},\ }\href {https://doi.org/10.1103/PhysRevC.77.041302} {\bibfield
  {journal} {\bibinfo  {journal} {Phys. Rev. C}\ }\textbf {\bibinfo {volume}
  {77}},\ \bibinfo {pages} {041302} (\bibinfo {year} {2008})}\BibitemShut
  {NoStop}%
\bibitem [{\citenamefont {Brodeur}\ \emph {et~al.}(2012)\citenamefont
  {Brodeur}, \citenamefont {Brunner}, \citenamefont {Champagne}, \citenamefont
  {Ettenauer}, \citenamefont {Smith}, \citenamefont {Lapierre}, \citenamefont
  {Ringle}, \citenamefont {Ryjkov}, \citenamefont {Bacca}, \citenamefont
  {Delheij}, \citenamefont {Drake}, \citenamefont {Lunney}, \citenamefont
  {Schwenk},\ and\ \citenamefont {Dilling}}]{BBC12}%
  \BibitemOpen
  \bibfield  {author} {\bibinfo {author} {\bibfnamefont {M.}~\bibnamefont
  {Brodeur}}, \bibinfo {author} {\bibfnamefont {T.}~\bibnamefont {Brunner}},
  \bibinfo {author} {\bibfnamefont {C.}~\bibnamefont {Champagne}}, \bibinfo
  {author} {\bibfnamefont {S.}~\bibnamefont {Ettenauer}}, \bibinfo {author}
  {\bibfnamefont {M.~J.}\ \bibnamefont {Smith}}, \bibinfo {author}
  {\bibfnamefont {A.}~\bibnamefont {Lapierre}}, \bibinfo {author}
  {\bibfnamefont {R.}~\bibnamefont {Ringle}}, \bibinfo {author} {\bibfnamefont
  {V.~L.}\ \bibnamefont {Ryjkov}}, \bibinfo {author} {\bibfnamefont
  {S.}~\bibnamefont {Bacca}}, \bibinfo {author} {\bibfnamefont
  {P.}~\bibnamefont {Delheij}}, \bibinfo {author} {\bibfnamefont {G.~W.~F.}\
  \bibnamefont {Drake}}, \bibinfo {author} {\bibfnamefont {D.}~\bibnamefont
  {Lunney}}, \bibinfo {author} {\bibfnamefont {A.}~\bibnamefont {Schwenk}},\
  and\ \bibinfo {author} {\bibfnamefont {J.}~\bibnamefont {Dilling}},\ }\href
  {https://doi.org/10.1103/PhysRevLett.108.052504} {\bibfield  {journal}
  {\bibinfo  {journal} {Phys. Rev. Lett.}\ }\textbf {\bibinfo {volume} {108}},\
  \bibinfo {pages} {052504} (\bibinfo {year} {2012})}\BibitemShut {NoStop}%
\bibitem [{\citenamefont {Sun}\ \emph {et~al.}(2021)\citenamefont {Sun},
  \citenamefont {Nakamura}, \citenamefont {Kondo}, \citenamefont {Satou},
  \citenamefont {Lee}, \citenamefont {Matsumoto}, \citenamefont {Ogata},
  \citenamefont {Kikuchi}, \citenamefont {Aoi}, \citenamefont {Ichikawa},
  \citenamefont {Ieki}, \citenamefont {Ishihara}, \citenamefont {Kobayshi},
  \citenamefont {Motobayashi}, \citenamefont {Otsu}, \citenamefont {Sakurai},
  \citenamefont {Shimamura}, \citenamefont {Shimoura}, \citenamefont
  {Shinohara}, \citenamefont {Sugimoto}, \citenamefont {Takeuchi},
  \citenamefont {Togano},\ and\ \citenamefont {Yoneda}}]{SNK21}%
  \BibitemOpen
  \bibfield  {author} {\bibinfo {author} {\bibfnamefont {Y.}~\bibnamefont
  {Sun}}, \bibinfo {author} {\bibfnamefont {T.}~\bibnamefont {Nakamura}},
  \bibinfo {author} {\bibfnamefont {Y.}~\bibnamefont {Kondo}}, \bibinfo
  {author} {\bibfnamefont {Y.}~\bibnamefont {Satou}}, \bibinfo {author}
  {\bibfnamefont {J.}~\bibnamefont {Lee}}, \bibinfo {author} {\bibfnamefont
  {T.}~\bibnamefont {Matsumoto}}, \bibinfo {author} {\bibfnamefont
  {K.}~\bibnamefont {Ogata}}, \bibinfo {author} {\bibfnamefont
  {Y.}~\bibnamefont {Kikuchi}}, \bibinfo {author} {\bibfnamefont
  {N.}~\bibnamefont {Aoi}}, \bibinfo {author} {\bibfnamefont {Y.}~\bibnamefont
  {Ichikawa}}, \bibinfo {author} {\bibfnamefont {K.}~\bibnamefont {Ieki}},
  \bibinfo {author} {\bibfnamefont {M.}~\bibnamefont {Ishihara}}, \bibinfo
  {author} {\bibfnamefont {T.}~\bibnamefont {Kobayshi}}, \bibinfo {author}
  {\bibfnamefont {T.}~\bibnamefont {Motobayashi}}, \bibinfo {author}
  {\bibfnamefont {H.}~\bibnamefont {Otsu}}, \bibinfo {author} {\bibfnamefont
  {H.}~\bibnamefont {Sakurai}}, \bibinfo {author} {\bibfnamefont
  {T.}~\bibnamefont {Shimamura}}, \bibinfo {author} {\bibfnamefont
  {S.}~\bibnamefont {Shimoura}}, \bibinfo {author} {\bibfnamefont
  {T.}~\bibnamefont {Shinohara}}, \bibinfo {author} {\bibfnamefont
  {T.}~\bibnamefont {Sugimoto}}, \bibinfo {author} {\bibfnamefont
  {S.}~\bibnamefont {Takeuchi}}, \bibinfo {author} {\bibfnamefont
  {Y.}~\bibnamefont {Togano}},\ and\ \bibinfo {author} {\bibfnamefont
  {K.}~\bibnamefont {Yoneda}},\ }\href
  {https://doi.org/https://doi.org/10.1016/j.physletb.2021.136072} {\bibfield
  {journal} {\bibinfo  {journal} {Phys. Lett. B}\ }\textbf {\bibinfo {volume}
  {814}},\ \bibinfo {pages} {136072} (\bibinfo {year} {2021})}\BibitemShut
  {NoStop}%
\bibitem [{\citenamefont {Aumann}\ \emph {et~al.}(1999)\citenamefont {Aumann},
  \citenamefont {Aleksandrov}, \citenamefont {Axelsson}, \citenamefont
  {Baumann}, \citenamefont {Borge}, \citenamefont {Chulkov}, \citenamefont
  {Cub}, \citenamefont {Dostal}, \citenamefont {Eberlein}, \citenamefont
  {Elze}, \citenamefont {Emling}, \citenamefont {Geissel}, \citenamefont
  {Goldberg}, \citenamefont {Golovkov}, \citenamefont {Gr\"unschlo\ss{}},
  \citenamefont {Hellstr\"om}, \citenamefont {Hencken}, \citenamefont
  {Holeczek}, \citenamefont {Holzmann}, \citenamefont {Jonson}, \citenamefont
  {Korshenninikov}, \citenamefont {Kratz}, \citenamefont {Kraus}, \citenamefont
  {Kulessa}, \citenamefont {Leifels}, \citenamefont {Leistenschneider},
  \citenamefont {Leth}, \citenamefont {Mukha}, \citenamefont {M\"unzenberg},
  \citenamefont {Nickel}, \citenamefont {Nilsson}, \citenamefont {Nyman},
  \citenamefont {Petersen}, \citenamefont {Pf\"utzner}, \citenamefont
  {Richter}, \citenamefont {Riisager}, \citenamefont {Scheidenberger},
  \citenamefont {Schrieder}, \citenamefont {Schwab}, \citenamefont {Simon},
  \citenamefont {Smedberg}, \citenamefont {Steiner}, \citenamefont {Stroth},
  \citenamefont {Surowiec}, \citenamefont {Suzuki}, \citenamefont {Tengblad},\
  and\ \citenamefont {Zhukov}}]{AAA99}%
  \BibitemOpen
  \bibfield  {author} {\bibinfo {author} {\bibfnamefont {T.}~\bibnamefont
  {Aumann}}, \bibinfo {author} {\bibfnamefont {D.}~\bibnamefont {Aleksandrov}},
  \bibinfo {author} {\bibfnamefont {L.}~\bibnamefont {Axelsson}}, \bibinfo
  {author} {\bibfnamefont {T.}~\bibnamefont {Baumann}}, \bibinfo {author}
  {\bibfnamefont {M.~J.~G.}\ \bibnamefont {Borge}}, \bibinfo {author}
  {\bibfnamefont {L.~V.}\ \bibnamefont {Chulkov}}, \bibinfo {author}
  {\bibfnamefont {J.}~\bibnamefont {Cub}}, \bibinfo {author} {\bibfnamefont
  {W.}~\bibnamefont {Dostal}}, \bibinfo {author} {\bibfnamefont
  {B.}~\bibnamefont {Eberlein}}, \bibinfo {author} {\bibfnamefont {T.~W.}\
  \bibnamefont {Elze}}, \bibinfo {author} {\bibfnamefont {H.}~\bibnamefont
  {Emling}}, \bibinfo {author} {\bibfnamefont {H.}~\bibnamefont {Geissel}},
  \bibinfo {author} {\bibfnamefont {V.~Z.}\ \bibnamefont {Goldberg}}, \bibinfo
  {author} {\bibfnamefont {M.}~\bibnamefont {Golovkov}}, \bibinfo {author}
  {\bibfnamefont {A.}~\bibnamefont {Gr\"unschlo\ss{}}}, \bibinfo {author}
  {\bibfnamefont {M.}~\bibnamefont {Hellstr\"om}}, \bibinfo {author}
  {\bibfnamefont {K.}~\bibnamefont {Hencken}}, \bibinfo {author} {\bibfnamefont
  {J.}~\bibnamefont {Holeczek}}, \bibinfo {author} {\bibfnamefont
  {R.}~\bibnamefont {Holzmann}}, \bibinfo {author} {\bibfnamefont
  {B.}~\bibnamefont {Jonson}}, \bibinfo {author} {\bibfnamefont {A.~A.}\
  \bibnamefont {Korshenninikov}}, \bibinfo {author} {\bibfnamefont {J.~V.}\
  \bibnamefont {Kratz}}, \bibinfo {author} {\bibfnamefont {G.}~\bibnamefont
  {Kraus}}, \bibinfo {author} {\bibfnamefont {R.}~\bibnamefont {Kulessa}},
  \bibinfo {author} {\bibfnamefont {Y.}~\bibnamefont {Leifels}}, \bibinfo
  {author} {\bibfnamefont {A.}~\bibnamefont {Leistenschneider}}, \bibinfo
  {author} {\bibfnamefont {T.}~\bibnamefont {Leth}}, \bibinfo {author}
  {\bibfnamefont {I.}~\bibnamefont {Mukha}}, \bibinfo {author} {\bibfnamefont
  {G.}~\bibnamefont {M\"unzenberg}}, \bibinfo {author} {\bibfnamefont
  {F.}~\bibnamefont {Nickel}}, \bibinfo {author} {\bibfnamefont
  {T.}~\bibnamefont {Nilsson}}, \bibinfo {author} {\bibfnamefont
  {G.}~\bibnamefont {Nyman}}, \bibinfo {author} {\bibfnamefont
  {B.}~\bibnamefont {Petersen}}, \bibinfo {author} {\bibfnamefont
  {M.}~\bibnamefont {Pf\"utzner}}, \bibinfo {author} {\bibfnamefont
  {A.}~\bibnamefont {Richter}}, \bibinfo {author} {\bibfnamefont
  {K.}~\bibnamefont {Riisager}}, \bibinfo {author} {\bibfnamefont
  {C.}~\bibnamefont {Scheidenberger}}, \bibinfo {author} {\bibfnamefont
  {G.}~\bibnamefont {Schrieder}}, \bibinfo {author} {\bibfnamefont
  {W.}~\bibnamefont {Schwab}}, \bibinfo {author} {\bibfnamefont
  {H.}~\bibnamefont {Simon}}, \bibinfo {author} {\bibfnamefont {M.~H.}\
  \bibnamefont {Smedberg}}, \bibinfo {author} {\bibfnamefont {M.}~\bibnamefont
  {Steiner}}, \bibinfo {author} {\bibfnamefont {J.}~\bibnamefont {Stroth}},
  \bibinfo {author} {\bibfnamefont {A.}~\bibnamefont {Surowiec}}, \bibinfo
  {author} {\bibfnamefont {T.}~\bibnamefont {Suzuki}}, \bibinfo {author}
  {\bibfnamefont {O.}~\bibnamefont {Tengblad}},\ and\ \bibinfo {author}
  {\bibfnamefont {M.~V.}\ \bibnamefont {Zhukov}},\ }\href
  {https://doi.org/10.1103/PhysRevC.59.1252} {\bibfield  {journal} {\bibinfo
  {journal} {Phys. Rev. C}\ }\textbf {\bibinfo {volume} {59}},\ \bibinfo
  {pages} {1252} (\bibinfo {year} {1999})}\BibitemShut {NoStop}%
\bibitem [{\citenamefont {Wang}\ \emph {et~al.}(2002)\citenamefont {Wang},
  \citenamefont {Galonsky}, \citenamefont {Kruse}, \citenamefont {Tryggestad},
  \citenamefont {White-Stevens}, \citenamefont {Zecher}, \citenamefont {Iwata},
  \citenamefont {Ieki}, \citenamefont {Horv\'ath}, \citenamefont {De\'ak},
  \citenamefont {Kiss}, \citenamefont {Seres}, \citenamefont {Kolata},
  \citenamefont {von Schwarzenberg}, \citenamefont {Warner},\ and\
  \citenamefont {Schelin}}]{WGK02}%
  \BibitemOpen
  \bibfield  {author} {\bibinfo {author} {\bibfnamefont {J.}~\bibnamefont
  {Wang}}, \bibinfo {author} {\bibfnamefont {A.}~\bibnamefont {Galonsky}},
  \bibinfo {author} {\bibfnamefont {J.~J.}\ \bibnamefont {Kruse}}, \bibinfo
  {author} {\bibfnamefont {E.}~\bibnamefont {Tryggestad}}, \bibinfo {author}
  {\bibfnamefont {R.~H.}\ \bibnamefont {White-Stevens}}, \bibinfo {author}
  {\bibfnamefont {P.~D.}\ \bibnamefont {Zecher}}, \bibinfo {author}
  {\bibfnamefont {Y.}~\bibnamefont {Iwata}}, \bibinfo {author} {\bibfnamefont
  {K.}~\bibnamefont {Ieki}}, \bibinfo {author} {\bibfnamefont {A.}~\bibnamefont
  {Horv\'ath}}, \bibinfo {author} {\bibfnamefont {F.}~\bibnamefont {De\'ak}},
  \bibinfo {author} {\bibfnamefont {A.}~\bibnamefont {Kiss}}, \bibinfo {author}
  {\bibfnamefont {Z.}~\bibnamefont {Seres}}, \bibinfo {author} {\bibfnamefont
  {J.~J.}\ \bibnamefont {Kolata}}, \bibinfo {author} {\bibfnamefont
  {J.}~\bibnamefont {von Schwarzenberg}}, \bibinfo {author} {\bibfnamefont
  {R.~E.}\ \bibnamefont {Warner}},\ and\ \bibinfo {author} {\bibfnamefont
  {H.}~\bibnamefont {Schelin}},\ }\href
  {https://doi.org/10.1103/PhysRevC.65.034306} {\bibfield  {journal} {\bibinfo
  {journal} {Phys. Rev. C}\ }\textbf {\bibinfo {volume} {65}},\ \bibinfo
  {pages} {034306} (\bibinfo {year} {2002})}\BibitemShut {NoStop}%
\bibitem [{\citenamefont {Bacca}\ \emph {et~al.}(2004)\citenamefont {Bacca},
  \citenamefont {Barnea}, \citenamefont {Leidemann},\ and\ \citenamefont
  {Orlandini}}]{BB04}%
  \BibitemOpen
  \bibfield  {author} {\bibinfo {author} {\bibfnamefont {S.}~\bibnamefont
  {Bacca}}, \bibinfo {author} {\bibfnamefont {N.}~\bibnamefont {Barnea}},
  \bibinfo {author} {\bibfnamefont {W.}~\bibnamefont {Leidemann}},\ and\
  \bibinfo {author} {\bibfnamefont {G.}~\bibnamefont {Orlandini}},\ }\href
  {https://doi.org/10.1103/PhysRevC.69.057001} {\bibfield  {journal} {\bibinfo
  {journal} {Phys. Rev. C}\ }\textbf {\bibinfo {volume} {69}},\ \bibinfo
  {pages} {057001} (\bibinfo {year} {2004})}\BibitemShut {NoStop}%
\bibitem [{\citenamefont {Pinilla}\ \emph {et~al.}(2012)\citenamefont
  {Pinilla}, \citenamefont {Descouvemont},\ and\ \citenamefont {Baye}}]{PDB12}%
  \BibitemOpen
  \bibfield  {author} {\bibinfo {author} {\bibfnamefont {E.~C.}\ \bibnamefont
  {Pinilla}}, \bibinfo {author} {\bibfnamefont {P.}~\bibnamefont
  {Descouvemont}},\ and\ \bibinfo {author} {\bibfnamefont {D.}~\bibnamefont
  {Baye}},\ }\href {https://doi.org/10.1103/PhysRevC.85.054610} {\bibfield
  {journal} {\bibinfo  {journal} {Phys. Rev. C}\ }\textbf {\bibinfo {volume}
  {85}},\ \bibinfo {pages} {054610} (\bibinfo {year} {2012})}\BibitemShut
  {NoStop}%
\end{thebibliography}%

\end{document}